\documentclass{article}

\usepackage[utf8]{inputenc}
\usepackage{authblk}
\usepackage[margin=0.9in,footskip=0.25in]{geometry}
\usepackage{amssymb}
\usepackage{latexsym}
\usepackage{url}
\usepackage{xcolor}
\usepackage{hyperref}
\definecolor{newcolor}{rgb}{.8,.349,.1}
\usepackage{graphicx}
\usepackage{upgreek}
\usepackage{bm}
\usepackage{amsmath,amsfonts}
\usepackage{diagbox}
\usepackage{caption}
\usepackage{subcaption}
\usepackage{tabularx, makecell}

\providecommand{\keywords}[1]{\textbf{\textit{Keywords: }} #1}
\DeclareMathSymbol{\shortminus}{\mathbin}{AMSa}{"39}
\setcellgapes{0.5pt}

\title{Rigid and non-rigid motion compensation in weight-bearing cone-beam CT of the knee using (noisy) inertial measurements}

\date{}

\author[1,2]{Jennifer~Maier\thanks{jennifer.maier@fau.de}}
\author[2]{Marlies~Nitschke}
\author[3]{Jang-Hwan~Choi}
\author[4]{Garry~Gold}
\author[5]{Rebecca~Fahrig}
\author[2]{Bjoern~M.~Eskofier}
\author[1]{Andreas~Maier}

\affil[1]{Pattern Recognition Lab, Friedrich-Alexander-Universit\"at Erlangen-N\"urnberg, Erlangen, Germany}
\affil[2]{Machine Learning and Data Analytics Lab, Friedrich-Alexander-Universit\"at Erlangen-N\"urnberg, Erlangen, Germany}
\affil[3]{Division of Mechanical and Biomedical Engineering, Ewha Womans University, Seoul, Republic of Korea}
\affil[4]{Department of Radiology, School of Medicine, Stanford University, Stanford, California, USA}
\affil[5]{Innovation, Advanced Therapies, Siemens Healthcare GmbH, Forchheim, Germany}

\begin{document}

\maketitle

\begin{abstract}
%%%
Involuntary subject motion is the main source of artifacts in weight-bearing cone-beam CT of the knee.
To achieve image quality for clinical diagnosis, the motion needs to be compensated. 
We propose to use inertial measurement units (IMUs) attached to the leg for motion estimation.
We perform a simulation study using real motion recorded with an optical tracking system.
Three IMU-based correction approaches are evaluated, namely rigid motion correction, non-rigid 2D projection deformation and non-rigid 3D dynamic reconstruction.
We present an initialization process based on the system geometry.
With an IMU noise simulation, we investigate the applicability of the proposed methods in real applications.
All proposed IMU-based approaches correct motion at least as good as a state-of-the-art marker-based approach.
The structural similarity index and the root mean squared error between motion-free and motion corrected volumes are improved by 24-35\% and 78-85\%, respectively, compared with the uncorrected case.
The noise analysis shows that the noise levels of commercially available IMUs need to be improved by a factor of $10^{5}$ which is currently only achieved by specialized hardware not robust enough for the application.
The presented study confirms the feasibility of this novel approach and defines improvements necessary for a real application.
%%%%
\end{abstract}


%% main text
\section{Introduction}
\label{sec:introduction}
\begin{figure}[tb]
\centering
    \subcaptionbox
        {\label{fig-intromodel_carm}}
        {\includegraphics[height=10pc]{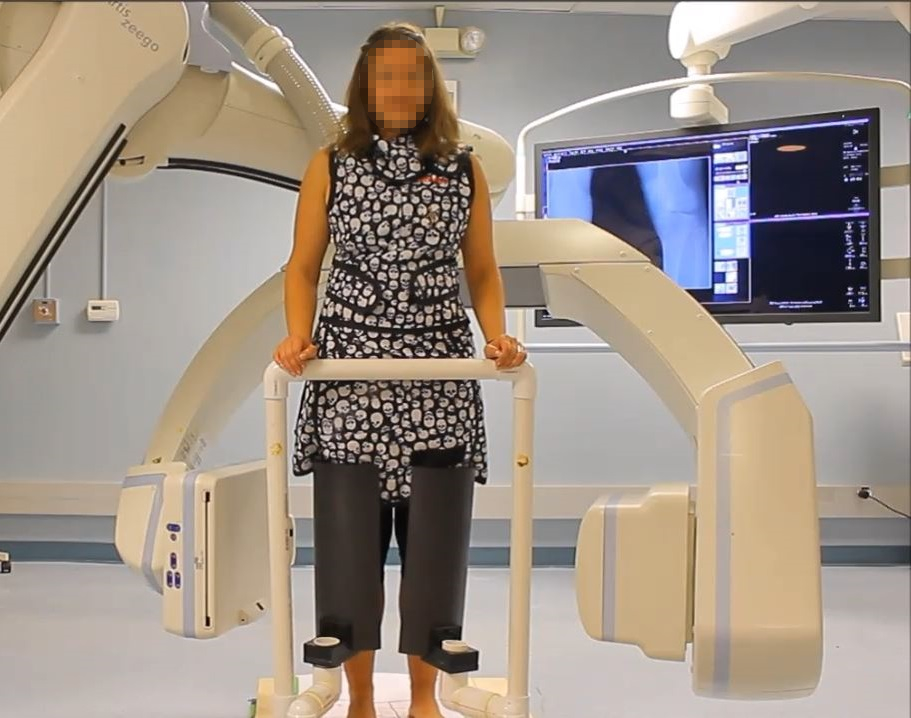}}
    %\hfill
    \quad
    \subcaptionbox
        {\label{fig-intromodel_model}}
        {\includegraphics[height=10pc]{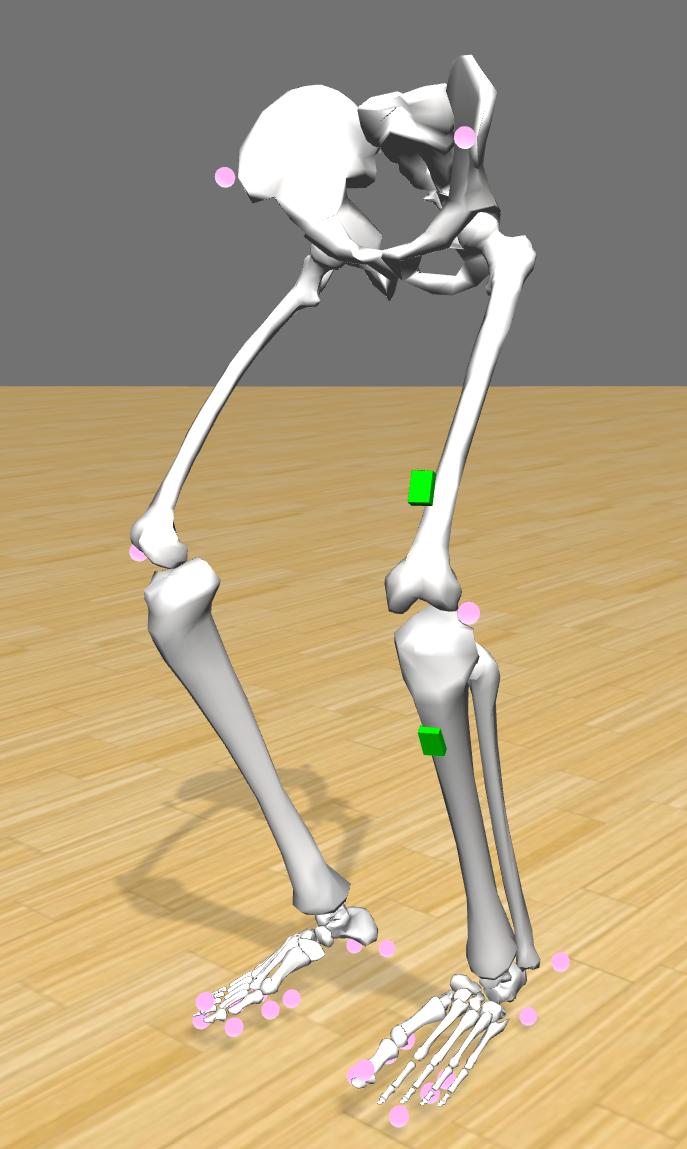}}
    \caption{(a) Setup of a weight-bearing C-arm cone-beam CT scan of the knees \cite{Maier2020}, (b) Biomechanical model with virtual reflective markers on the legs (pink spheres) and IMUs on thigh and shin (green boxes).}
    \label{fig-intromodel}
\end{figure}

Patients suffering from osteoarthritis (OA) experience joint pain due to an increased porosity of articular cartilage or even a complete loss of tissue \cite{Arden2006}.
This pain is especially severe in the frequently loaded knee joint.
Since cartilage degeneration also affects the mechanical properties of the tissue, an analysis of the behavior under load can help to understand and analyze diseased tissue \cite{Powers2003}.
Contrast-agent enhanced CT-imaging in a weight-bearing standing position is an established method to visualize the articular cartilage in the knee joint \cite{Choi2013,Choi2014}.
It is realized by using a flexible robotic C-arm system scanning on a horizontal trajectory around the patient's knees \cite{Maier2011}.
The setup of such a scan can be seen in Fig.~\ref{fig-intromodel_carm}.
One drawback of this setting is that subjects tend to have a higher range of motion when standing compared with the supine lying position, which leads to image artifacts like blurring and double edges in the resultant reconstructions.
Motion corrupted reconstructions lose their diagnostic value and are unsuitable for further processing.
Since it is not useful to prevent subject motion when aiming for natural stance, the movement during the scan has to be estimated and corrected.

Previous approaches are either image-based or use an external signal or marker in order to correct for motion.
Performing 2D/3D registration showed very good motion compensation capabilities, but required prior bone segmentations and is computationally expensive \cite{Berger2016}.
The same limitation holds for an approach based on a penalized image sharpness criterion \cite{Sisniega2017}.
By leveraging the epipolar consistency of CT scans, the translation but not the rotation of the knees during a CT scan was estimated \cite{Bier2017}. Bier et al. \cite{Bier2018landmark} proposed to estimate motion by tracking anatomical landmarks in the projection images using a neural network.
Until now, their approach was not applied for motion compensation and was only reliable if there were no other objects present.
An investigation on the practicality of using range cameras for motion compensated reconstruction showed promising results on simulated data \cite{Bier2018range}. 
An established and effective method for motion compensation in weight-bearing imaging of the knee is based on small metallic markers attached to the leg and tracked in the projection images to iteratively estimate 3D motion  \cite{Choi2013,Choi2014}.
However, the process of placing the markers is tedious and they produce metal artifacts in areas of interest in the resulting images.

In C-arm CT, inertial measurement units (IMUs) containing an accelerometer and a gyroscope have until now been applied for navigation \cite{Jost2016} and calibration \cite{Lemammer2019} purposes.
Our recent work was the first to propose the use of IMUs for motion compensation in weight-bearing cone-beam CT (CBCT) of the knee \cite{Maier2020}.
We evaluated the feasibility of using the measurements of one IMU placed on the shin of the subject for rigid motion compensation in a simulation study.
However, since the actual movement during the scan is non-rigid, not all artifacts could be resolved with the rigid correction approach.
For this reason, we now investigate non-rigid motion compensation based on 2D or 3D deformation using signals recorded by two IMUs placed on the shin and the thigh.
Furthermore, a method to estimate the initial pose and velocity of the sensors is presented.
These two parameters are needed for motion estimation and were assumed to be known in the aforementioned publication \cite{Maier2020}.
Another drawback of our previous publication is that we only simulated optimal IMU signals and neglected possible measurement errors.
In order to assess the applicability of our proposed methods in a real setting, and as a third contribution, we now analyze how sensor noise added to the optimal IMU signals influences the motion compensation capabilities.

In this article, we present a simulation study similar to the one in our previous publication, therefore some content of section \ref{sec-simulation} is closely related to Maier et al. \cite{Maier2020}.
Furthermore, the previously published rigid motion estimation approach is repeated for better comprehensibility.
%Since this article is an extension of \cite{Maier2020}, some parts of section \ref{sec-methods} are adapted from that publication.
%The content is repeated for better comprehensibility.
%

\section{Materials and methods}
\label{sec-methods} 
\begin{figure}[tb]
  \centerline{\includegraphics[width=25pc]{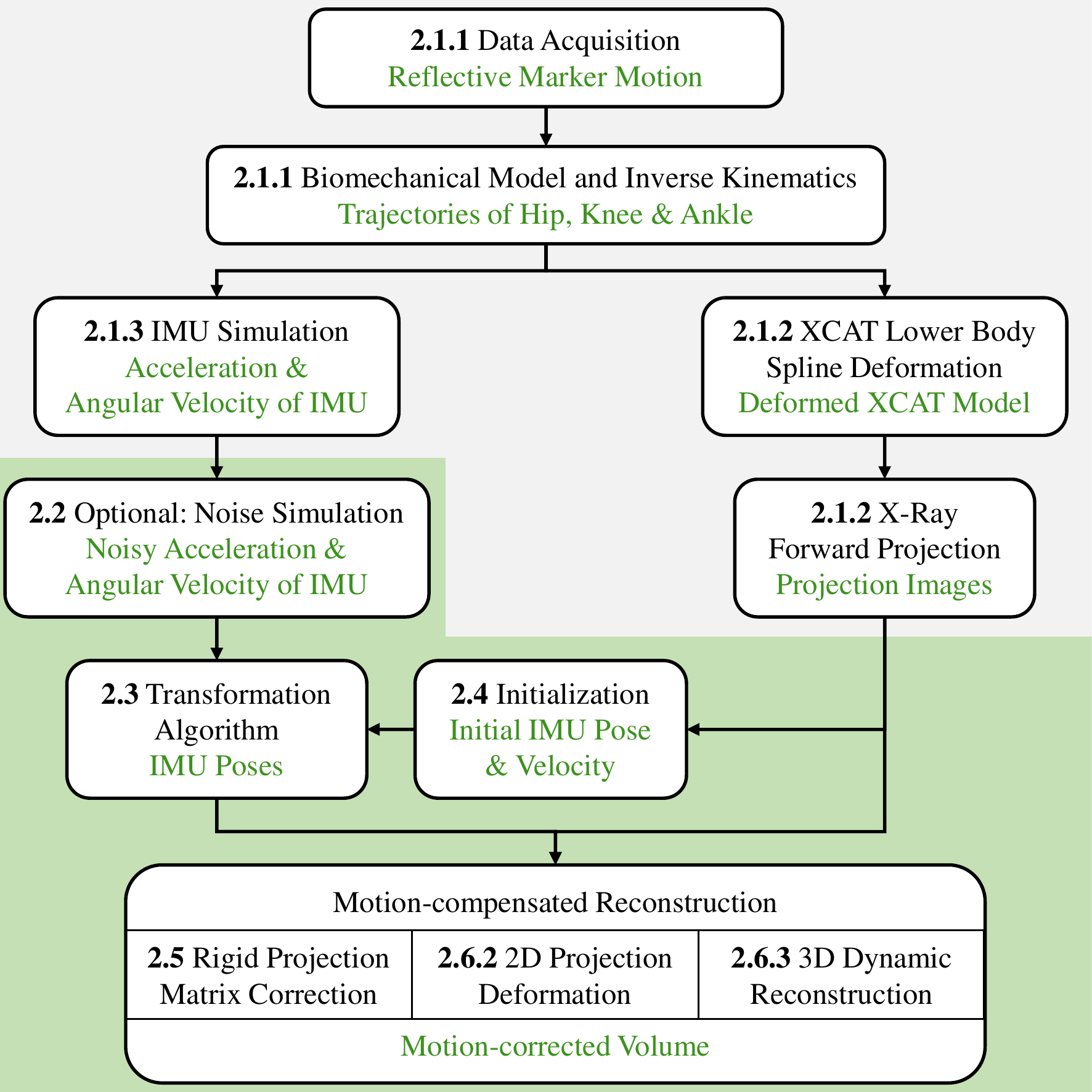}}
  \caption{Processing Pipeline presented in section \ref{sec-methods}. Black font: processing steps, green font: respective output. The simulation study is presented in section \ref{sec-simulation} (shaded in grey). Sections \ref{sec-noise} to \ref{sec-nonrigid} (shaded in green) describe the proposed data processing.}
  \label{fig:graph}
\end{figure}
The whole processing pipeline of the presented simulation study is shown in Fig.~\ref{fig:graph}, where black font describes each processing step and green font the respective output.
All steps shaded in gray relate to the simulation and are described in Section \ref{sec-simulation}, while all steps shaded in green refer to the proposed data processing presented in Sections \ref{sec-noise} to \ref{sec-nonrigid}.

The simulation contains the following steps:
The motion of standing subjects is recorded with an optical motion capture system and used to animate a biomechanical model to obtain the trajectories of hip, knees and ankles (\ref{sec-Biomecsimulation}).
These positions are then used in two ways:
First, the lower body of a numerical phantom is deformed to mimic the subject motion and a motion-corrupted C-arm CBCT scan is simulated (\ref{sec-XCATsimulation}).
Secondly, the signals of IMUs placed on the model's leg are computed (\ref{sec-IMUsimulation}).
%The single steps are explained in more detail in the indicated sections.

In Section \ref{sec-noise}, measurement noise is added to the optimal sensor signals.
These noisy signals are later used to analyze the influence of measurement errors on the motion correction with IMUs.

%Subject motion is recorded with an optical motion capture system.
%The recorded motion is used to animate a biomechanical model and by inverse kinematics computation the moving positions of hip, knees and ankles are obtained.
%These positions are then used in two ways.
%First, the lower body of the numerical XCAT model is deformed to mimic the subject motion, and projection images of a simulated C-arm CBCT scan are created from the XCAT model deformed at each time frame.
%Second, the measurements of IMUs placed on the model's leg are computed.

Then, the proposed IMU-based approaches for motion compensated reconstruction of the motion-corrupted CT scan are described:
From the IMU measurements, the position and orientation, i.e. the pose, of the IMUs over time are computed (\ref{sec-poseestimation}).
For this step, the initial sensor pose and velocity need to be known and are estimated from the first two projection images (\ref{sec-initialization}).
The computed poses are then used for three different motion correction approaches compared in this article.
First, rigid motion matrices are computed from the IMU poses and used to adapt the projection matrices for reconstruction (\ref{sec-rigid}).
Second, the projection images are non-rigidly deformed before 3D reconstruction (\ref{sec-nonrigid2D}).
Third, the sensor poses are incorporated in the reconstruction algorithm for a 3D non-rigid deformation (\ref{sec-nonrigid3D}).

%In section \ref{sec-noise}, measurement noise is added to the optimal sensor signals to analyze the influence of measurement errors on the pose estimation and motion correction with IMUs.

\subsection{Simulation}
\label{sec-simulation}
\subsubsection{Data acquisition and biomechanical model}
\label{sec-Biomecsimulation}
In order to create realistic simulations, real motion of standing persons is acquired.
Seven healthy subjects are recorded in three settings of 20 seconds duration: holding a squat of 30 degrees and 60 degrees knee flexion, and actively performing squats.
Seven reflective markers are attached to each subject's sacrum, to the right and left anterior superior iliac spine, to the right and left lateral epicondyle of the knee, and to the right and left malleolus lateralis.
The marker positions are tracked with a 3D optical motion tracking system (Vicon, Oxford, UK) at a sampling rate of 120\,Hz.

Subsequently, in the software OpenSim \cite{Delp2007}, the marker positions of the active squatting scan of each subject are used to scale a biomechanical model of the human lower body \cite{Hamner2010} to the subject's anthropometry.
The model with attached virtual markers shown in pink is displayed in Fig.~\ref{fig-intromodel_model}.

The scaled model is then animated two times per subject by computing the inverse kinematics using the marker positions of the 30 degrees and 60 degrees squatting scans \cite{Seth2018}.
%\cite{Buss2004}.
The inverse kinematics computation results in the generalized coordinates that best represent the measured motion.
These generalized coordinates describe the complete model motion as the global position and orientation of the pelvis and the angles of all leg joints.
Before further processing, jumps in the data that occur due to noise are removed, and the signals are filtered with a second order Butterworth filter with a cutoff frequency of 6\,Hz in order to remove system noise.
Since the model is scaled to the subject's anthropometry, the generalized coordinates can be used to compute the trajectories of hip, knee and ankle joints.
These joint trajectories are the input to all further steps of the data simulation.

\subsubsection{XCAT deformation and CT projection generation}
\label{sec-XCATsimulation}
We generate a virtual motion-corrupted CT scan using the moving 4D extended cardiac-torso (XCAT) phantom \cite{Segars2010}.
The legs of the numerical phantom consist of the bones tibia, fibula, femur and patella including bone marrow and surrounded by body soft tissue.
All structures contained in the phantom have material-specific properties.
Their shapes are defined by 3D control points spanning non-uniform rational B-splines.
By changing the positions of these control points the structures of the XCAT phantom can be non-rigidly deformed.

In the default XCAT phantom the legs are extended.
To simulate a standing C-arm CT scan, the phantom needs to take on the squatting pose of the recorded subjects that is varying over time.
For this purpose, the positions of the XCAT spline control points are changed based on the hip, knee, and ankle positions of the biomechanical model.
The deformation process is described in detail by Choi et al. \cite{Choi2013}.

Then, a horizontal circular CT scan of the knees is simulated.
As in a real setting, 248\,projections are generated with an angular increment of 0.8\,degrees between projections, corresponding to a sampling rate of 31\,Hz.
The virtual C-arm rotates on a trajectory with 1198\,mm source detector distance and 780\,mm source isocenter distance.
The detector has a size of $620\times480$ pixels with an isotropic pixel resolution of 0.616\,mm.
In a natural standing position, the knees are too far apart to both fit on the detector, therefore the rotation center is placed in the center of the left leg of the deformed XCAT phantom.
Then, forward projections are created as described in Maier et al. \cite{Maier2012}.
Since the subject of this study is to analyze the motion compensation capability of IMUs, CBCT artifacts other than motion are not included in the simulation.

\subsubsection{Simulation of IMU measurements}
\label{sec-IMUsimulation}
The trajectories of hip, knees and ankles computed using the biomechanical model are used to simulate the measurements of IMUs placed on the leg of the model.
IMUs are commonly used for motion analysis in sports and movement disorders \cite{Kautz2017}.
They are low cost, small and lightweight devices that measure their acceleration and angular velocity on three perpendicular axes.
Besides the motion signal, the accelerometer measures the earth's gravitational field distributed on its three axes depending on its orientation.
We virtually place two such sensors on the shin, 14\,cm below the left knee joint and on the thigh, 25\,cm below the hip joint aligned with the respective body segment (Fig.~\ref{fig-intromodel_model}).
In a future real application, sensors in these positions are visible in the projections as needed for initialization (\ref{sec-initialization}).
At the same time, they are situated at a sufficient distance from the knee joint in the direction of the CBCT rotation axis such that their metal components do not cause artifacts in the region of interest.

The simulated acceleration $\mathbf{a}(t)$ and angular velocity $\bm{\omega}(t)$ at time point $t$ are computed as described by \cite{Bogert1996,Desapio2017}:
\begin{equation}
    \mathbf{a}(t) = \mathbf{R}(t)^\top(\ddot{\mathbf{r}}_{Seg}(t) + \ddot{\mathbf{R}}(t) \mathbf{p}_{Sen}(t) - \mathbf{g})\,,
\end{equation}
\begin{equation}
    [\bm{\omega}(t)]_{\times} = \mathbf{R}(t)^\top\dot{\mathbf{R}}(t) = 
    \begin{pmatrix}
        0 & -\omega_{z}(t) & \omega_{y}(t) \\
        \omega_{z}(t) & 0 & -\omega_{x}(t) \\
        -\omega_{y}(t) & \omega_{x}(t) & 0
    \end{pmatrix}\,,
\end{equation}
\begin{equation}
    \bm{\omega}(t) = (\omega_{x}(t), \omega_{y}(t), \omega_{z}(t))^\top\,.
\end{equation}
All parameters required in these equations are obtained by performing forward kinematics of the biomechanical model.
The 3$\times$3 rotation matrix $\mathbf{R}(t)$ describes the orientation of the sensor at time point $t$ in the global coordinate system, $\dot{\mathbf{R}}(t)$ and $\ddot{\mathbf{R}}(t)$ are its first and second order derivatives with respect to time.
The position of the segment the sensor is mounted on in the global coordinate system at time point $t$ is described by $\mathbf{r}_{Seg}(t)$, with $\ddot{\mathbf{r}}_{Seg}(t)$ being its second order derivative.
$\mathbf{p}_{Sen}(t)$ is the position of the sensor in the local coordinate system of the segment the sensor was mounted on.
Parameter $\mathbf{g}=(0,-9.80665, 0)^\top$ is the global gravity vector.
\subsection{IMU noise simulation}
\label{sec-noise}
The IMU signal computation assumes a perfect IMU that can measure without the influence of errors.
However, in a real application errors can have a significant influence preventing an effective motion compensation.
For example, Kok et al. \cite{Kok2017} showed that the integration of a stationary IMU over ten seconds leads to errors in the orientation and position estimates of multiple degrees and meters, respectively.

The most prominent error sources in IMUs leading to these high deviations are random measurement noise and an almost constant bias \cite{Woodman2007}.
In this study, we focus on the analysis of the unpredictable sensor noise.
Commercially available consumer IMU devices have noise densities that are acceptable for larger motion analysis.
An example of a commercially available sensor BMI160, Bosch Sensortec GmbH, Reutlingen, Germany, has an output noise density of $180\,{\mu}g/s^2$ and $0.007\,^\circ$/s and a root mean square (RMS) noise of $1.8\,mg/s^2$ and $0.07\,^\circ$/s at $200\,Hz$ \cite{Bosch2020}.
However, our data shows that the signals produced by a standing swaying motion have amplitudes in the range of $0.3\,mg/s^2$ resp. $0.02\,^\circ$/s.
This means that when measuring with an off-the-shelf sensor, the signal would be completely masked by noise.
%%Such noise levels can only be reached by highly sensitive specialized sensors that are far too delicate for the application at hand \cite{Yamane2019}. 
For this reason, we investigate the noise level improvement necessary to use IMUs for the task of motion compensation in weight-bearing CT imaging.

We simulate white Gaussian noise signals of different RMS levels and add them onto the simulated acceleration $\mathbf{a}(t)$ and angular velocity $\bm{\omega}(t)$.
Starting with the RMS values of the aforementioned Bosch sensor, the noise level is divided by factors of ten down to a factor of $10^{5}$.
The accelerometer and gyroscope noise levels are decreased independently.
In the following, we will use the notation $f_a$ and $f_g$ for the exponent, i.e. the factor the RMS value is divided by is $10^{f_a}$ resp. $10^{f_g}$.
The noisy IMU signals are then used to compute rigid motion matrices for motion compensation as explained in Sections \ref{sec-poseestimation} and \ref{sec-rigid}.

Note that the noise influence is evaluated independently of the IMU-based motion compensation methods.
All motion compensation methods presented in the following sections are first evaluated on the noise-free signals.
Afterwards, we perform rigid motion compensation with noisy IMU signals to investigate the influence of noise on the applicability of IMUs for motion compensation.
\subsection{Transformation algorithm}
\label{sec-poseestimation}
The following descriptions are based on Maier et al. \cite{Maier2020} and are required for all IMU-based motion compensation approaches presented in this article.

The IMU measures motion in its local coordinate system, however, motion in the global coordinate system of the CBCT scan is required for motion compensation.
The orientation and position of the IMU $\mathbf{S}(t)$ in the global coordinate system at each frame $t$ is described by the affine matrix
\begin{equation}
    \mathbf{S}(t) =
    \begin{pmatrix}
        \begin{array}{c|c}
              \mathbf{R}(t) & \mathbf{r}(t) \\ 
              \hline
              \mathbf{0}^\top & 1
        \end{array}
    \end{pmatrix}\,,
\end{equation}
where $\mathbf{R}(t)$ is a 3$\times$3 rotation matrix, $\mathbf{r}(t)$ is a 3$\times$1 translation vector, and $\mathbf{0}$ is the 3$\times$1 zero-vector.
The IMU pose can be updated for each subsequent frame using the affine global pose change matrix $\bm{\Updelta}_{g}(t)$:
\begin{equation}
    \label{eq:Siplus1}
    \mathbf{S}(t+1) = \bm{\Updelta}_{g}(t)\mathbf{S}(t)\,.
\end{equation}
This global change $\bm{\Updelta}_{g}(t)$ can be computed by transforming the local change in the IMU coordinate system $\bm{\Updelta}_{l}(t)$ to the global coordinate system using the current IMU pose:
\begin{equation}
    \label{eq:deltaglobal}
    \bm{\Updelta}_{g}(t) = \mathbf{S}(t)\bm{\Updelta}_{l}(t)\mathbf{S}(t)^{-1}\,.
\end{equation}
Thus, if the initial pose $\mathbf{S}(t=0)$ is known, the problem is reduced to estimating the local pose change $\bm{\Updelta}_{l}(t)$ in the IMU coordinate system, which is described in the following paragraphs.

The gyroscope measures the orientation change over time $\bm{\omega}(t)$ on the three axes of the IMU's local coordinate system which can be directly used to rotate the IMU from frame to frame.
The measured acceleration $\mathbf{a}(t)$, however, needs to be processed to obtain the position change over time.
First, the gravity measured on the IMU's three axes is removed based on its global orientation.
For this purpose, the angular velocity $\bm{\omega}(t)$ is rewritten to 3$\times$3 rotation matrices $\mathbf{G}(t)$ and used to update the global orientation of the sensor $\mathbf{R}(t)$.
This orientation can then be used to obtain the gravity vector $\mathbf{g}(t)$ in the local coordinate system for each frame $t$:
\begin{align}
    \mathbf{R}(t+1) &= \mathbf{R}(t) \mathbf{G}(t)\,,\\
    \mathbf{g}(t) &= \mathbf{R}(t)^\top \mathbf{g}\,.
\end{align}
The sensor's local velocity $\mathbf{v}(t)$, i.e. its position change over time, is then computed as the integral of the gravity-free acceleration considering the sensor's orientation changes:
\begin{equation}
    \label{eq:velocity}
    \mathbf{v}(t+1) = \mathbf{G}(t)^\top(\mathbf{a}(t) + \mathbf{g}(t) + \mathbf{v}(t))\,.
\end{equation}

With these computations, the desired local pose change of the IMU $\bm{\Updelta}_{l}(t)$ for each frame $t$ is expressed as an affine matrix containing the local rotation change and position change:
\begin{equation}
    \label{eq:deltalocal}
    \bm{\Updelta}_{l}(t) =
    \begin{pmatrix}
        \begin{array}{c|c}
              \mathbf{G}(t) & \mathbf{v}(t) \\ 
              \hline
              \mathbf{0}^\top & 1
        \end{array}
    \end{pmatrix}\,.
\end{equation}
Note that the initial pose $\mathbf{S}(t=0)$ and velocity $\mathbf{v}(t=0)$ need to be known or estimated in order to apply this transformation process.

\subsection{IMU pose and velocity initialization}
\label{sec-initialization}
\begin{figure}[tb]
    \centering
    \subcaptionbox
        {\label{fig-initialization_pose}}
        {\includegraphics[height=10pc]{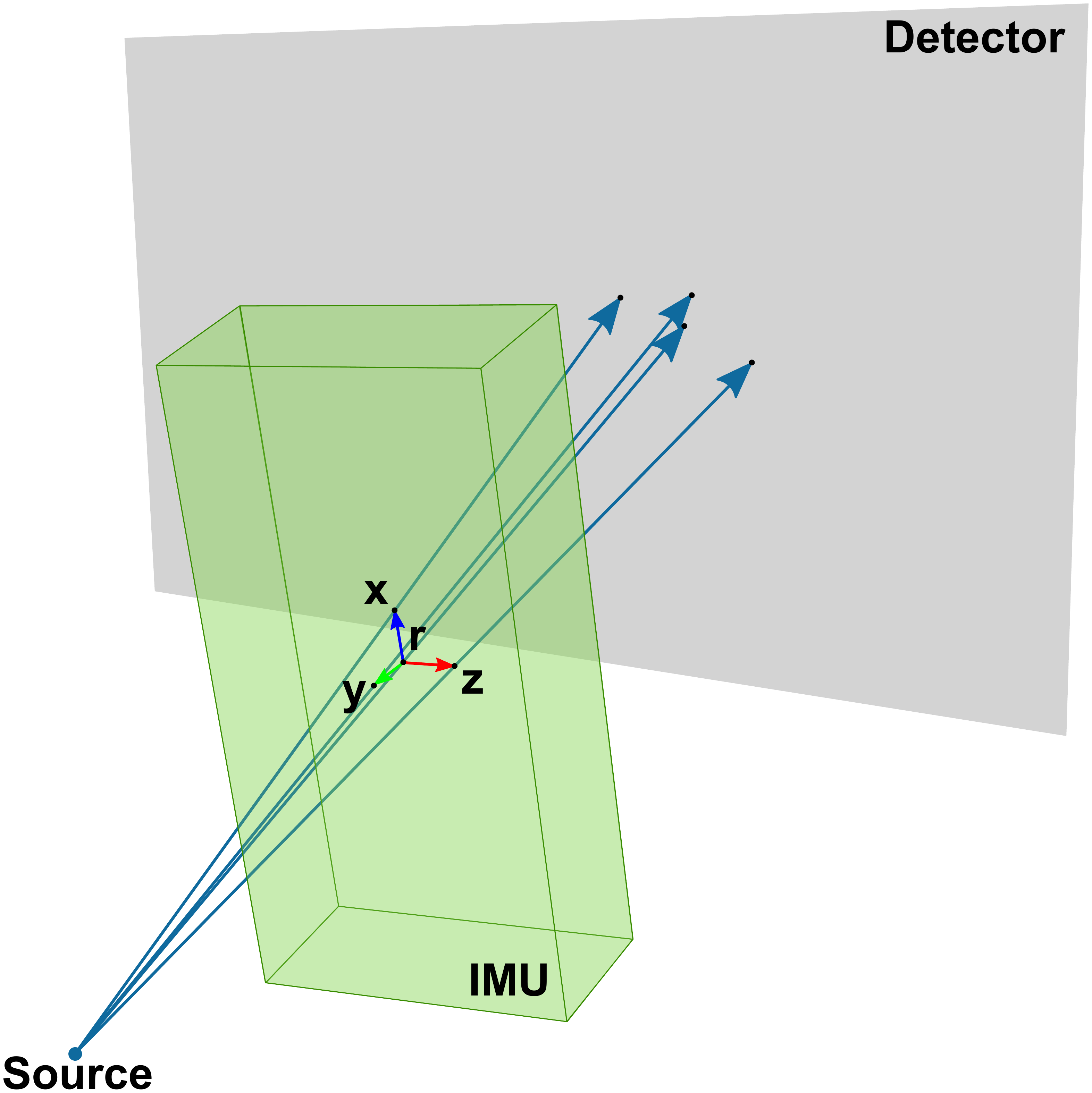}}
    %\hfill
    \quad
    \subcaptionbox
        {\label{fig-initialization_velocity}}
        {\includegraphics[height=7pc]{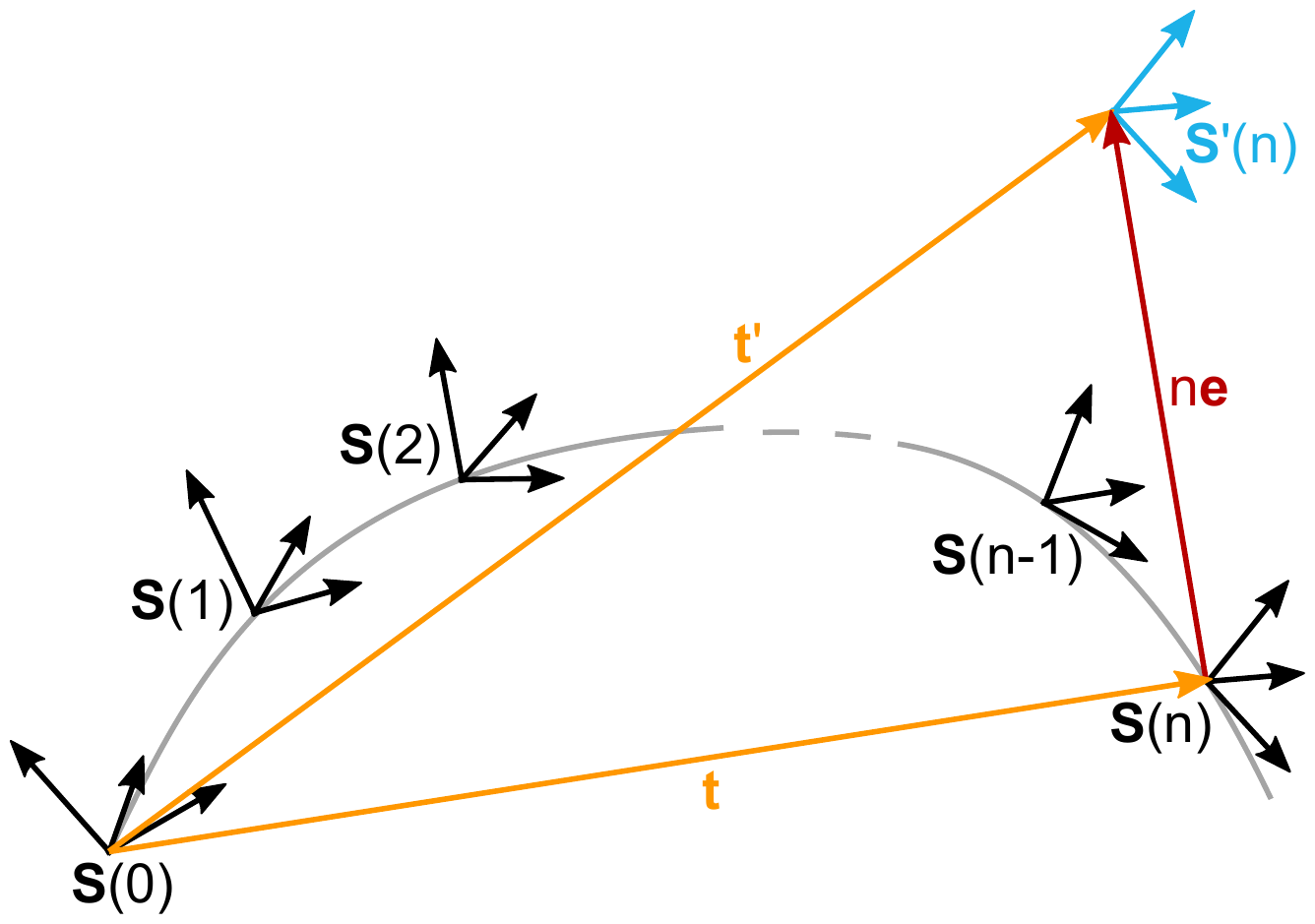}}
    \caption{(a) Disproportionate visualization of the initialization concept. The green box represents the sensor with its coordinate system plotted inside. The X-rays (blue) pass through the metal components and hit the detector (gray). (b) Visualization of the velocity initialization approach. Computing the pose $\mathbf{S}'(t=n)$ with incorrect initial velocity $\mathbf{v}(t=0)$ leads to a wrong translation $\mathbf{t}'$ which is used for velocity initialization.}
    \label{fig-initialization}
\end{figure}
In our previously published work, the initial pose and velocity of the IMU necessary for pose estimation in (\ref{eq:Siplus1}) and (\ref{eq:velocity}) were assumed to be known, which is not the case in a real setting \cite{Maier2020}.
Thies et al. \cite{Thies2019} proposed to estimate the initial pose as an average sensor pose computed from the complete set of projection images.
However, using the average position over the multi-second scan including subject motion leads to inaccurate motion compensation results.
For this reason, we present an initial pose estimation based only on the first projection image.
By incorporating also the second projection image, the initial velocity can be estimated.

\subsubsection{Initial IMU pose}
\label{sec-initialization1}
The pose of the IMU $\mathbf{S}(t)$ at frame $t$ contains the 3D position of the origin $\mathbf{r}(t)$ and the three perpendicular axes of measurement $\mathbf{u_x}(t)$, $\mathbf{u_y}(t)$ and $\mathbf{u_z}(t)$ in the rotation matrix $\mathbf{R}(t)$.
This coordinate system can be computed from any four non-coplanar points within the IMU if their geometrical relation to the sensor's measurement coordinate system is known.
For simplicity, in this simulation study we assume that these four points are the IMU's origin $\mathbf{r}$ and the points $\mathbf{x}$, $\mathbf{y}$ and $\mathbf{z}$ at the tips of the three axes' unit vectors $\mathbf{u_x}$, $\mathbf{u_y}$, and $\mathbf{u_z}$.
We also assume that the sensor has small, highly attenuating metal components at these four points making their projected 2D positions easy to track on an X-ray projection image.
Since the C-arm system geometry is calibrated prior to performing CT acquisitions, the 3D position of each 2D detector pixel, and the 3D source position for each projection are also known.
Then the searched points $\mathbf{r}(t)$, $\mathbf{x}(t)$, $\mathbf{y}(t)$, and $\mathbf{z}(t)$ are positioned on the straight line between the source and the respective projected point (Fig.~\ref{fig-initialization_pose}).

In the considered case, the four points need to fulfill the following properties:
\begin{itemize}
    \item The vectors $\mathbf{u_x}(t)$, $\mathbf{u_y}(t)$ and $\mathbf{u_z}(t)$ spanned by $\mathbf{r}(t)$, $\mathbf{x}(t)$, $\mathbf{y}(t)$ and $\mathbf{z}(t)$ must have unit length.
    \item The euclidean distance between two of the points $\mathbf{x}(t)$, $\mathbf{y}(t)$ and $\mathbf{z}(t)$ must be $\sqrt{2}$.
    \item The inner product of two of the vectors $\mathbf{u_x}(t)$, $\mathbf{u_y}(t)$ and $\mathbf{u_z}(t)$ must be zero.
    \item The right-handed cross product of two of the vectors $\mathbf{u_x}(t)$, $\mathbf{u_y}(t)$ and $\mathbf{u_z}(t)$ must result in the third vector.
\end{itemize}
Solving the resulting non-linear system of equations defined by these constraints for the first projection at time point $t=0$ yields the 3D positions of $\mathbf{r}(t=0)$, $\mathbf{x}(t=0)$, $\mathbf{y}(t=0)$, and $\mathbf{z}(t=0)$ and thereby the initial sensor pose $\mathbf{S}(t=0)$.

\subsubsection{Initial IMU velocity}
\label{sec-initialization2}
The initial IMU velocity $\mathbf{v}(t=0)$ is needed to compute the velocity update in (\ref{eq:velocity}).
In the following paragraphs, we describe a process to estimate the initial velocity, which is illustrated in Fig.~\ref{fig-initialization_velocity}.

The IMU acquires data with a higher sampling rate than the C-arm acquires projection images (120\,Hz and 31\,Hz, respectively).
If the two systems are synchronized the correspondence between the sampling time points $t$ of the IMU and the projection image acquisition points $i$ is known.
The first projection image corresponds to the first IMU sample at time point $t=i=0$ and is used to estimate the initial pose $\mathbf{S}(t=0)$.
The second projection image at $i=1$ corresponds to the IMU sampling point $t=n$ with $n>1$ and is used to estimate the pose $\mathbf{S}(t=n)$.

Since each IMU pose can be computed from the previous one by applying the pose change between frames with (\ref{eq:Siplus1}) and (\ref{eq:deltaglobal}), the IMU pose at frame $t=n$ can also be expressed as
\begin{equation}
    \mathbf{S}(t=n) = \mathbf{S}(t=0)\bm{\Updelta}_{l}(t=0)\bm{\Updelta}_{l}(t=1)\ldots\bm{\Updelta}_{l}(t=n-1)\,,
\end{equation}
which can be rearranged to
\begin{equation}
    \mathbf{S}(t=0)^{-1}\mathbf{S}(t=n) = \bm{\Updelta}_{l}(t=0)\bm{\Updelta}_{l}(t=1)\ldots\bm{\Updelta}_{l}(t=n-1)\,.
\end{equation}
However, since $\mathbf{v}(t=0)$ is not known, also $\bm{\Updelta}_{l}(t=0)$ and all subsequent local change matrices are not known.
Therefore, instead of the actual $\mathbf{v}(t=0)$, we use the zero-vector as initial velocity introducing an error vector $\mathbf{e}$:
\begin{equation}
    \mathbf{v}'(t=0) = \mathbf{0} = \mathbf{v}(t=0) + \mathbf{e}\,.
\end{equation}
This error is propagated and accumulated in the frame by frame velocity computation in (\ref{eq:velocity}) and for $t>=1$ the resulting error-prone velocity is
\begin{equation}
    \label{eq:velocityprime}
    \mathbf{v}'(t) = \mathbf{v}(t) + \mathbf{G}(t-1)^\top\mathbf{G}(t-2)^\top\ldots\mathbf{G}(0)^\top\mathbf{e}\,.
\end{equation}
These error-prone velocities $\mathbf{v}'(t)$ lead to incorrect pose change matrices $\bm{\Updelta}_{l}'(t)$ and thereby to an incorrect computation of $\mathbf{S}'(t=n)$:
\begin{align}
    \bm{\Updelta}'_{l}(t) &=
    \begin{pmatrix}
        \begin{array}{c|c}
              \mathbf{G}(t) & \mathbf{v}'(t) \\ 
              \hline
              \mathbf{0}^\top & 1
        \end{array}
    \end{pmatrix}\,,\\
    \mathbf{S}(t=0)^{-1}\mathbf{S}'(t=n) &= \bm{\Updelta}'_{l}(t=0)\bm{\Updelta}'_{l}(t=1)\ldots\bm{\Updelta}'_{l}(t=n-1)\,.
    \label{eq:Sprime}
\end{align}
Inserting (\ref{eq:velocityprime}) and expanding (\ref{eq:Sprime}) shows that the incorrect initial velocity only has an effect on the translation of the resulting affine matrix:
\begin{equation}
    \mathbf{S}(t=0)^{-1}\mathbf{S}'(t=n) = \mathbf{S}(t=0)^{-1}\mathbf{S}(t=n)+
    \begin{pmatrix}
        \begin{array}{c|c}
              \mathbf{0}_{3,3} & n\mathbf{e} \\ 
              \hline
              \mathbf{0}^\top & 1
        \end{array}
    \end{pmatrix}\,.
\end{equation}
In this equation, $\mathbf{0}_{3,3}$ denotes a 3$\times$3 matrix filled with zeros.
%
%The translation \mathbf{t} of $\mathbf{S}(t=0)^{-1}\mathbf{S}(t=n)$ is computed using the poses estimated from the projection images and the translation $\mathbf{t}'$ of $\mathbf{S}(t=0)^{-1}\mathbf{S}'(t=n)$ is computed using the estimated initial pose and the incorrect initial velocity $\mathbf{v}'(t=0)$, leading to:
If the translation of $\mathbf{S}(t=0)^{-1}\mathbf{S}'(t=n)$ is denoted as $\mathbf{t}'$ and the translation of $\mathbf{S}(t=0)^{-1}\mathbf{S}(t=n)$ is denoted as $\mathbf{t}$, this leads to:
\begin{equation}
    \mathbf{t}' = \mathbf{t} + n\mathbf{e}.
\end{equation}
The correct initial velocity $\mathbf{v}(t=0)$ is computed as
\begin{equation}
    \mathbf{v}(t=0) = -\mathbf{e} = -\frac{1}{n}\cdot(\mathbf{t}' - \mathbf{t})\,.
\end{equation}

\subsection{Rigid projection matrix correction}
\label{sec-rigid}
Under the assumption that the legs move rigidly during the CT scan, it is sufficient to use the measurements of only one sensor placed e.g. on the shin for motion estimation.
The pose change matrices estimated in (\ref{eq:deltaglobal}) and (\ref{eq:deltalocal}) can then be directly applied for motion correction.
Note that the angular velocity and velocity are resampled to the CT acquisition frequency before pose change computation using the synchronized correspondences between C-arm and IMU.

An affine motion matrix $\mathbf{M}(i)$ containing the rotation and translation is computed for each projection $i$.
The motion matrix for the first projection $i=0$ is defined as the identity matrix $\mathbf{M}(i=0)=\mathbf{I}$, i.e. the pose at the first projection is used as the reference pose.
Each subsequent matrix is then obtained using the global pose change matrix computed from the sensor measurements:
\begin{equation}
    \mathbf{M}(i+1) = \mathbf{M}(i) \bm{\Updelta}_{g}(i)\,.
\end{equation}

In order to correct for the motion during the CBCT scan, we then modify the projection matrices $\mathbf{P}(i)$ of the calibrated CT scan with the motion matrices $\mathbf{M}(i)$ resulting in motion corrected projection matrices $\hat{\mathbf{P}}(i)$:
\begin{equation}
    \hat{\mathbf{P}}(i) = \mathbf{P}(i) \mathbf{M}(i)\,.
\end{equation}
The corrected projection matrices are then used for the volume reconstruction as described in Section \ref{sec-evaluationMoCo}.

\subsection{Non-rigid motion correction}
\label{sec-nonrigid}
Contrary to the assumption in Section \ref{sec-rigid}, the leg motion during the scan is non-rigid since the subjects are not able to hold exactly the same squatting angle for the duration of the scan.
As a consequence, the motion can not entirely be described by a rigid transformation.
To address this issue, we propose a non-rigid motion correction using both IMUs placed on the model.
Using the formulas presented in \ref{sec-poseestimation}, we can compute the poses $^t\mathbf{S}(t)$ and $^f\mathbf{S}(t)$ of the IMUs on tibia and femur, respectively.
Since the placement of the IMUs on the segments relative to the joints is known, the IMU poses can be used to describe the positions of ankle, knee and hip joint, $\mathbf{a}(t)$, $\mathbf{k}(t)$ and $\mathbf{h}(t)$, at each time point $t$.
%The estimated initial 3D positions of hip, knee and ankle at $t=0$ are stored as $\mathbf{h}(0)$, $\mathbf{k}(0)$ and $\mathbf{a}(0)$.
%For each next time step $t$ corresponding to a projection image, the 3D positions $\mathbf{h}(t)$, $\mathbf{k}(t)$ and $\mathbf{a}(t)$ of hip, knee and ankle are estimated using the sensor poses $^t\mathbf{S}(t)$ and $^f\mathbf{S}(t)$.

These estimated joint positions are used to non-rigidly correct for motion during the scans.
We propose two approaches that make use of Moving Least Squares (MLS) deformations in order to correct for motion \cite{Schaefer2006,Zhu2007}.
The first approach applies a 2D deformation to each projection image, and the second approach performs a 3D dynamic reconstruction where the deformation is integrated into the volume reconstruction.

\subsubsection{Moving least squares deformation}
The idea of MLS deformation is that the deformation of a scene is defined by a set of $m$ control points.
The original positions of the control points are denoted as $\mathbf{p}_j$, and their deformed positions are $\mathbf{q}_j$ with ${j=1,...,m}$.
For each pixel $\bm{\nu}$ in the image or volume, the goal is to find its position in the deformed image or volume depending on these control points.
For this purpose, the affine transformation $f(\bm{\nu})$ that minimizes the weighted distance between the known and estimated deformed positions should be found:
\begin{equation}
    \label{eq:mls}
    \sum_{j} \omega_j \lvert{f(\mathbf{p}_j)-\mathbf{q}_j}\rvert^2\,.
\end{equation}
This optimization is performed for each pixel individually, since the weights $\omega_j$ depend on the distance of the pixel $\bm{\nu}$ to the control points $\mathbf{p}_j$:
\begin{equation}
    \omega_j = \frac{1}{\lvert{\mathbf{p}_j-\bm{\nu}}\rvert^2}\,.
\end{equation}
The weighted centroids $\mathbf{p}_*$ and $\mathbf{q}_*$ and the shifted control points $\mathbf{\hat{p}}_j = \mathbf{p}_j - \mathbf{p}_*$ and $\mathbf{\hat{q}}_j = \mathbf{q}_j - \mathbf{q}_*$ are used in order to find the optimal solution of (\ref{eq:mls}) in both the 2D and 3D case:
\begin{align} 
    \mathbf{p}_* &= \frac{\sum_j\omega_j\mathbf{p}_j}{\sum_j\omega_j}\,, \\
    \mathbf{q}_* &= \frac{\sum_j\omega_j\mathbf{q}_j}{\sum_j\omega_j}\,.
\end{align}

According to Sch\"afer et al. \cite{Schaefer2006}, in the 2D image deformation case, the transformation minimizing (\ref{eq:mls}) is described by:
\begin{equation}
    \label{eq-2D}
    f(\bm{\nu}) = \lvert\bm{\nu}-\mathbf{p}_*\rvert \frac{\sum_j\mathbf{\hat{q}}_j\mathbf{A_j}}{\lvert\sum_j\mathbf{\hat{q}}_j\mathbf{A_j}\rvert} + \mathbf{q}_*\,,
\end{equation}
where
\begin{equation}
    \mathbf{A}_j = \omega_j
    \begin{pmatrix}
      \mathbf{\hat{p}}_j\\
      -\mathbf{\hat{p}}_j^\bot\\
    \end{pmatrix}
    \begin{pmatrix}
      \bm{\nu} - \mathbf{p}_*\\
      -(\bm{\nu} - \mathbf{p}_*)^\bot\\
    \end{pmatrix}\,.
\end{equation}

Finding the transformation that minimizes (\ref{eq:mls}) in the 3D case requires the computation of a singular value decomposition, as explained by Zhu et al. \cite{Zhu2007}:
\begin{equation}
    \sum_j \omega_j\hat{\mathbf{p}}_j\hat{\mathbf{q}}_j^T = \mathbf{U\Sigma{V}}^T\,.
\end{equation}
%If the determinant of $\mathbf{VU}^T$ is negative, the last column of $\mathbf{V}$ has to be negated before further processing \cite{Arun1987}.
The optimal transformation is then described by: 
\begin{equation}
    \label{eq-3D}
    f(\bm{\nu}) = \mathbf{VU}^T(\bm{\nu}-\mathbf{p}_*) + \mathbf{q}_*\,.
\end{equation}

\subsubsection{2D projection deformation}
\label{sec-nonrigid2D}
In our first proposed non-rigid approach, we deform the content of the 2D projection images in order to correct for motion.
The initial pose of the subject is used as reference pose, so the first projection image $i=0$ is left unaltered.
Each following projection image acquired at time point $i$ is transformed by MLS deformation using the estimated hip, knee and ankle joint positions $\mathbf{h}(i)$, $\mathbf{k}(i)$ and $\mathbf{a}(i)$ by using them as control points for the MLS deformation as described in the following paragraph.

To obtain the 2D points needed for a 2D projection image deformation, the 3D positions $\mathbf{h}(i)$, $\mathbf{k}(i)$ and $\mathbf{a}(i)$ are forward projected onto the detector using the system geometry.
However, since the detector is too small to cover the whole leg of a subject, the projected positions of the hip and ankle would be outside of the detector area.
For this reason, 3D points situated closer to the knee on the straight line between hip and knee, and on the straight line between ankle and knee are computed with $\alpha = 0.8$:
\begin{align}
    \label{eq-closertoknee1}
    \mathbf{h}'(i) &= (1-\alpha)\mathbf{h}(i) + \alpha{\mathbf{k}(i)}\,, \\
    \label{eq-closertoknee2}
    \mathbf{a}'(i) &= (1-\alpha)\mathbf{a}(i) + \alpha{\mathbf{k}(i)}\,.
\end{align}
Then, for each projection $i$, the initial 3D reference positions $\mathbf{a}'(i=0)$, $\mathbf{k}(i=0)$ and $\mathbf{h}'(i=0)$ are forward projected onto the detector resulting in the 2D control points $\mathbf{p}_j(i)$ with ${j = 1,2,3}$.
The 3D positions $\mathbf{h}'(i)$, $\mathbf{k}(i)$ and $\mathbf{a}'(i)$ at time of projection acquisition $i$ are forward projected to obtain $\mathbf{q}_j(i)$ with ${j = 1,2,3}$.
Each projection image is then deformed by computing the transformation $f(\bm{\nu})$ according to (\ref{eq-2D}) for each image pixel using these control points.
Finally, the motion corrected 3D volume is reconstructed from the resulting deformed projection images as described in Section \ref{sec-evaluationMoCo}.

\subsubsection{3D dynamic reconstruction}
\label{sec-nonrigid3D}
The second proposed non-rigid approach applies 3D deformations during volume reconstruction.
In the typical back-projection process of CT reconstruction, the 3D position of each voxel of the output volume is forward projected onto the detector for each projection image $i$, and the value at the projected position is added to the 3D voxel value.
For the proposed 3D dynamic reconstruction, this process is altered:
before forward projecting the 3D voxel position onto the detector for readout, it is transformed using 3D MLS deformation.
However, the readout value is added at the original voxel position.

For MLS deformation during reconstruction, the estimated positions of hip, knee and ankle joint $\mathbf{h}(i)$, $\mathbf{k}(i)$ and $\mathbf{a}(i)$ are used.
The reference pose is again the first pose at $i=0$ and the 3D positions of hip, knee and ankle $\mathbf{h}(i=0)$, $\mathbf{k}(i=0)$ and $\mathbf{a}(i=0)$ are used as control points $\mathbf{p}_j$ with ${j = 1,2,3}$.
The 3D positions $\mathbf{h}(i)$, $\mathbf{k}(i)$ and $\mathbf{a}(i)$ are used as $\mathbf{q}_j(i)$ with ${j = 1,2,3}$.
Note that the 3D positions $\mathbf{p}_j$ are the same for each projection, contrary to the 2D approach where they depend on forward projection using the system geometry.
Using these control points, during reconstruction the transformation $f(\bm{\nu})$ is computed for each voxel of the output volume and each projection according to (\ref{eq-3D}) and applied for deformation resulting in a motion-compensated output volume.

\section{Evaluation}
\subsection{IMU-based motion compensation}
\label{sec-evaluationMoCo}
All volumes are reconstructed by GPU accelerated filtered back-projection in the software framework CONRAD \cite{Maier2013}.
The filtered back-projection pipeline included a cosine weighting, a Parker weighting, a truncation correction and a Shepp Logan ramp filtering.
The reconstructed volumes have a size of $512^3$\,voxels with isotropic spacing of 0.5\,mm.
In the case of rigid motion compensation, the motion compensated projection matrices $\mathbf{P}'$ are used for reconstruction.
In the case of 2D non-rigid motion compensation, the deformed projection images are reconstructed using the original projection matrices.
In the case of 3D non-rigid motion compensation, the original projection matrices and projection images are used, but the back-projection process is adapted as described in Section \ref{sec-nonrigid3D}.

For comparison, an uncorrected motion-corrupted volume is reconstructed.
Furthermore, a motion-free reference is realized by simulating a CT scan where the initial pose of the model is kept constant throughout the scan.
The IMU-based motion compensation approaches are compared with a marker-based gold standard approach \cite{Choi2014}.
For this purpose, small highly attenuating metal markers placed on the knee joint are added to the projections and tracked for motion compensation as proposed by Choi et al. \cite{Choi2014}.
All volumes are scaled from 0 to 1 and registered to the motion-free reference reconstruction.

The image quality is compared against the motion-free reference by the structural similarity index measure (SSIM) and the root mean squared error (RMSE).
The SSIM index ranges from 0 (no similarity) to 1 (identical images) and considers differences in luminance, contrast and structure \cite{Wang2004}.
The metrics are computed on the whole reconstructed leg and on the lower leg and upper leg separately.

\subsection{Noise analysis}
The influence of noise on the motion correction is evaluated in two ways:

First, the estimated motion is compared by decomposing the motion matrices resulting from the noise-free signal and from the different levels of noisy signals into three-axial translations and rotations.
Each noisy result is then compared to the noise-free estimate.
For comparison, we compute the RMSE between each axis of the noise-free and the noisy translations and rotations, and then average over the three axes.
We only evaluate on one scan of one subject, but average over five independent repetitions of adding random white noise and computing motion matrices and RMSE.
%The resulting average RMSE is more conclusive than the result for one single run.

Secondly, volumes reconstructed from noisy signal motion estimates are analyzed.
Based on the RMSE results from the first part of the analysis, certain noise levels are chosen for rigid motion compensated reconstruction.
Rigid motion matrices are computed from the noisy signals as described in Sections \ref{sec-poseestimation} and \ref{sec-rigid} and used for volume reconstruction as described above.
For image quality comparison, the SSIM and RMSE are again computed.

\section{Results}
\subsection{IMU-based motion compensation}
The proposed initialization method yields the correct initial pose and velocity for all scans and all further computations are based on these estimates.
\begin{figure*}[tp]
    \centering
    \subcaptionbox*{}{\includegraphics[width=0.14\textwidth]{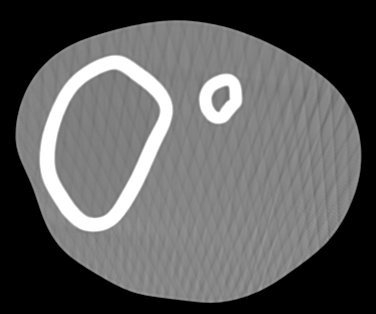}}
    \quad
    \subcaptionbox*{}{\includegraphics[width=0.14\textwidth]{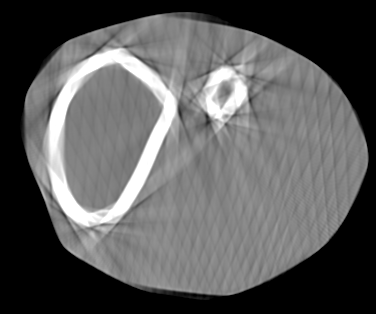}}
    \quad
    \subcaptionbox*{}{\includegraphics[width=0.14\textwidth]{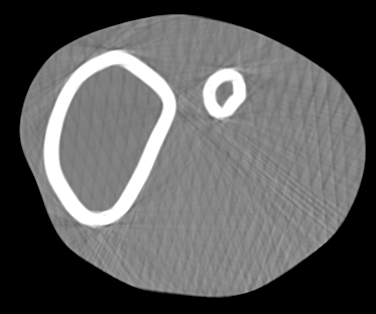}}
    \quad
    \subcaptionbox*{}{\includegraphics[width=0.14\textwidth]{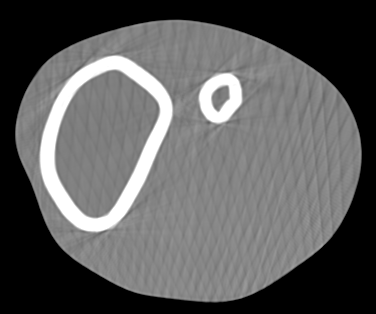}}
    \quad
    \subcaptionbox*{}{\includegraphics[width=0.14\textwidth]{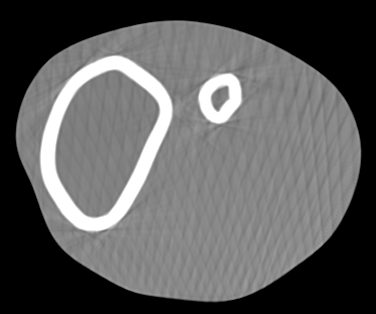}}
    \quad
    \subcaptionbox*{}{\includegraphics[width=0.14\textwidth]{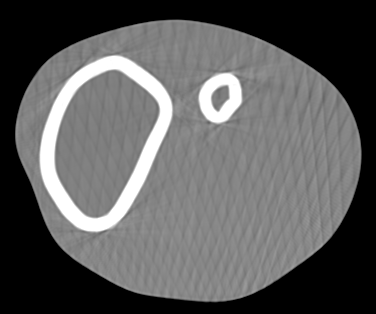}}
    \\
    \subcaptionbox*{}{\includegraphics[width=0.14\textwidth]{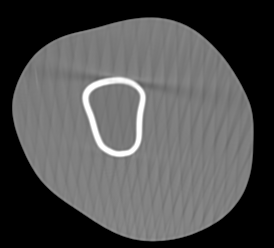}}
    \quad
    \subcaptionbox*{}{\includegraphics[width=0.14\textwidth]{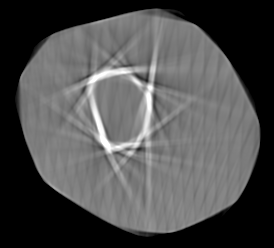}}
    \quad
    \subcaptionbox*{}{\includegraphics[width=0.14\textwidth]{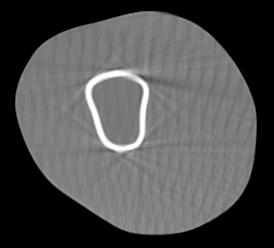}}
    \quad
    \subcaptionbox*{}{\includegraphics[width=0.14\textwidth]{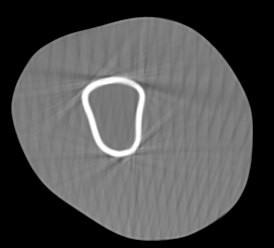}}
    \quad
    \subcaptionbox*{}{\includegraphics[width=0.14\textwidth]{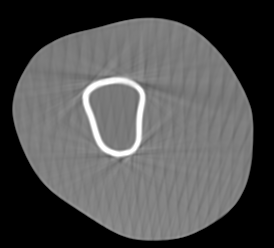}}
    \quad
    \subcaptionbox*{}{\includegraphics[width=0.14\textwidth]{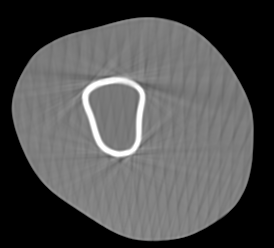}}
    \\
    \subcaptionbox{No motion}{\includegraphics[width=0.14\textwidth]{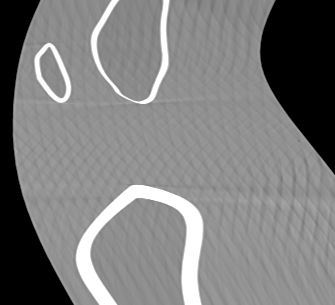}}
    \quad
    \subcaptionbox{Uncorrected}{\includegraphics[width=0.14\textwidth]{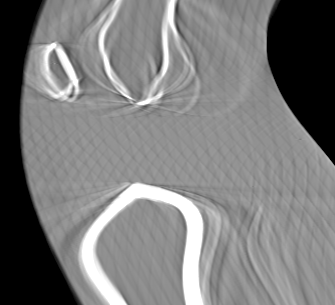}}
    \quad
    \subcaptionbox{Marker-based}{\includegraphics[width=0.14\textwidth]{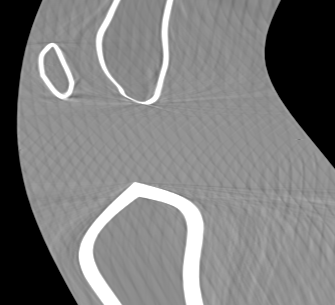}}
    \quad
    \subcaptionbox{Rigid}{\includegraphics[width=0.14\textwidth]{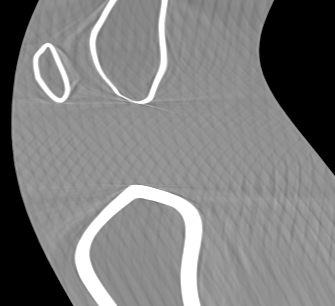}}
    \quad
    \subcaptionbox{2D non-rigid}{\includegraphics[width=0.14\textwidth]{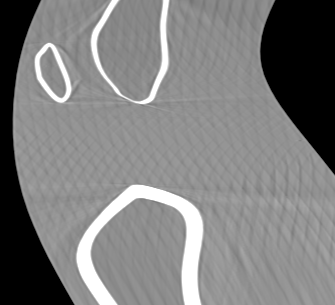}}
    \quad
    \subcaptionbox{3D non-rigid}{\includegraphics[width=0.14\textwidth]{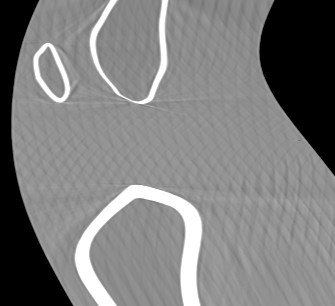}}
    \caption{Exemplary slices of a reconstructed volume. Rows: axial slice through shin, axial slice through thigh, sagittal slice. (a) Scan without motion, (b) uncorrected case, (c)  marker-based reference method, (d) rigid IMU method, (e) non-rigid IMU 2D projection deformation, (f) non-rigid IMU 3D dynamic reconstruction. Motion artifacts can be reduced by all proposed methods in a similar manner as the marker-based method.}
    \label{fig-result}
\end{figure*}
Figure~\ref{fig-result} shows axial slices through the tibia and the femur, and a sagittal slice of one example reconstruction.
All proposed methods are able to compensate for motion equally as well as the marker-based reference approach, or even slightly better.
\begin{figure}[tb]
    \centering
    \subcaptionbox*{}{\includegraphics[width=6pc]{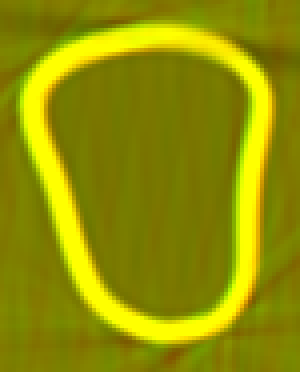}}
    \quad
    \subcaptionbox*{}{\includegraphics[width=6pc]{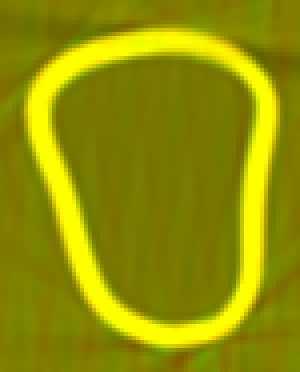}}
    \quad
    \subcaptionbox*{}{\includegraphics[width=6pc]{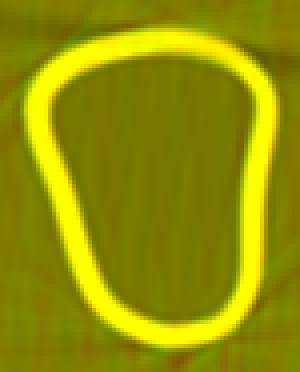}}
    \\
    \subcaptionbox{Rigid}{\includegraphics[width=6pc]{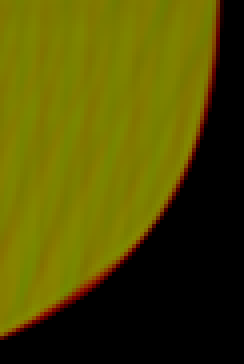}}
    \quad
    \subcaptionbox{2D non-rigid}{\includegraphics[width=6pc]{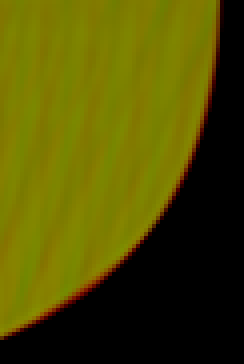}}
    \quad
    \subcaptionbox{3D non-rigid}{\includegraphics[width=6pc]{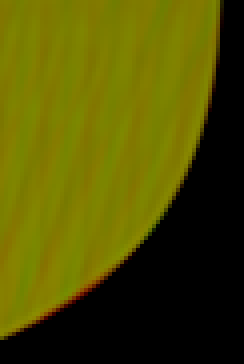}}
    \caption{Details of an axial slice through the thigh. Rows: femoral bone, skin border. Overlay of motion-free reference (red) and the result of the method (a) rigid IMU, (b) non-rigid 2D IMU, (c) non-rigid 3D IMU (green), overlapping pixels shown in yellow.}
    \label{fig-resultzoom}
\end{figure}
Differences between the methods can only be seen in a detailed overlay of the motion compensated reconstruction with the motion-free reconstruction.
In Fig.~\ref{fig-resultzoom}, details of the axial slice through the thigh are depicted at the femoral bone and at the skin-air-border.
The motion-free reconstruction is shown in red, and the motion compensated reconstructions of the rigid, non-rigid 2D and non-rigid 3D IMU methods are shown in green in the three columns.
All pixels that occur in both overlaid images are depicted in yellow.
It is noticeable that the rigid correction method fails to estimate the exact thigh motion leading to an observable shift as a red or green halo at the bone interface and at the skin border.
This is reduced for the non-rigid 2D correction and almost imperceptible for the non-rigid 3D correction.
\begin{table*}[tp]
\caption{Average SSIM and RMSE values with standard deviation of the motion compensated reconstructions compared to the motion-free reconstruction, where all volumes were scaled from 0 to 1. 30 degrees and 60 degrees squats are evaluated separately in the topmost two blocks, while the last block shows the average results over all scans. All measures were computed on the whole reconstructed part of the leg, and on the shank resp. thigh only.}
\label{tab-results}
\begin{small}
\begin{tabular*}{\textwidth}{@{\extracolsep{\fill}}l|lll|lll}
\hline
                 
				 & \multicolumn{3}{c|}{SSIM}                                                              & \multicolumn{3}{c}{RMSE}                                                              \\ \hline
                 & \multicolumn{1}{c}{Whole Leg} & \multicolumn{1}{c}{Shank} & \multicolumn{1}{c|}{Thigh} & \multicolumn{1}{c}{Whole Leg} & \multicolumn{1}{c}{Shank} & \multicolumn{1}{c}{Thigh} \\ \hline
                 & \multicolumn{6}{c}{30 degrees squat} \\
				 \hline
Uncorrected      & 0.777 $\pm$ 0.089             & 0.799 $\pm$ 0.093         & 0.742 $\pm$ 0.084         & 0.084 $\pm$ 0.029             & 0.081 $\pm$ 0.030         & 0.088 $\pm$ 0.029         \\
Marker-based           & 0.986 $\pm$ 0.002             & 0.986 $\pm$ 0.002         & 0.984 $\pm$ 0.003         & 0.021 $\pm$ 0.006             & 0.021 $\pm$ 0.006         & 0.022 $\pm$ 0.005         \\
Rigid IMU        & 0.993 $\pm$ 0.003             & 0.993 $\pm$ 0.004         & 0.992 $\pm$ 0.004         & 0.015 $\pm$ 0.005             & 0.015 $\pm$ 0.005         & 0.015 $\pm$ 0.005         \\
Non-rigid IMU 2D & 0.991 $\pm$ 0.003             & 0.991 $\pm$ 0.003         & 0.992 $\pm$ 0.002         & 0.017 $\pm$ 0.005             & 0.018 $\pm$ 0.005         & 0.016 $\pm$ 0.005         \\
Non-rigid IMU 3D & 0.994 $\pm$ 0.002             & 0.994 $\pm$ 0.003         & 0.994 $\pm$ 0.002         & 0.014 $\pm$ 0.004             & 0.014 $\pm$ 0.004         & 0.013 $\pm$ 0.004         \\ \hline
& \multicolumn{6}{c}{60 degrees squat} \\ \hline
Uncorrected      & 0.777 $\pm$ 0.111             & 0.790 $\pm$ 0.090         & 0.736 $\pm$ 0.118         & 0.079 $\pm$ 0.022             & 0.076 $\pm$ 0.020         & 0.084 $\pm$ 0.025         \\
Marker-based           & 0.982 $\pm$ 0.009             & 0.979 $\pm$ 0.012         & 0.974 $\pm$ 0.013         & 0.026 $\pm$ 0.009             & 0.025 $\pm$ 0.009         & 0.027 $\pm$ 0.009         \\
Rigid IMU        & 0.991 $\pm$ 0.006             & 0.990 $\pm$ 0.004         & 0.987 $\pm$ 0.006         & 0.018 $\pm$ 0.006             & 0.018 $\pm$ 0.007         & 0.019 $\pm$ 0.006         \\
Non-rigid IMU 2D & 0.989 $\pm$ 0.005             & 0.989 $\pm$ 0.005         & 0.989 $\pm$ 0.005         & 0.018 $\pm$ 0.006             & 0.018 $\pm$ 0.006         & 0.018 $\pm$ 0.007         \\
Non-rigid IMU 3D & 0.994 $\pm$ 0.004             & 0.991 $\pm$ 0.004         & 0.991 $\pm$ 0.004         & 0.016 $\pm$ 0.006             & 0.017 $\pm$ 0.006         & 0.016 $\pm$ 0.006         \\
\hline
& \multicolumn{6}{c}{All scans} \\ \hline
Uncorrected      & 0.774 $\pm$ 0.092             & 0.794 $\pm$ 0.090         & 0.739 $\pm$ 0.098         & 0.081 $\pm$ 0.025             & 0.078 $\pm$ 0.025         & 0.086 $\pm$ 0.026         \\
Marker-based           & 0.981 $\pm$ 0.009             & 0.982 $\pm$ 0.009         & 0.979 $\pm$ 0.011         & 0.024 $\pm$ 0.007             & 0.023 $\pm$ 0.007         & 0.024 $\pm$ 0.008         \\
Rigid IMU        & 0.991 $\pm$ 0.004             & 0.991 $\pm$ 0.004         & 0.990 $\pm$ 0.006         & 0.017 $\pm$ 0.006             & 0.016 $\pm$ 0.006         & 0.017 $\pm$ 0.006         \\
Non-rigid IMU 2D & 0.990 $\pm$ 0.004             & 0.990 $\pm$ 0.004         & 0.990 $\pm$ 0.004         & 0.017 $\pm$ 0.006             & 0.018 $\pm$ 0.006         & 0.017 $\pm$ 0.006         \\
Non-rigid IMU 3D & 0.993 $\pm$ 0.004             & 0.992 $\pm$ 0.004         & 0.993 $\pm$ 0.004         & 0.015 $\pm$ 0.005             & 0.015 $\pm$ 0.005         & 0.014 $\pm$ 0.005         \\
\hline
\end{tabular*}
\end{small}
\end{table*}
This visual impression is confirmed by the SSIM and RMSE values in Table~\ref{tab-results}.
All proposed methods achieve SSIM and RMSE values that are similar or better than those of the reference marker-based method.
Compared with the uncorrected case, this denotes an improvement of 24-35\% in the SSIM and 78-85\% in the RMSE values, respectively.
Higher SSIM scores and lower RMSE values are achieved for the 30 degrees squat scans compared with the 60 degrees squat scans.
When comparing the three proposed IMU methods, the results show a slight advantage of the non-rigid 3D approach over the other two IMU-based approaches.

\subsection{Noise analysis}
\begin{table*}[tp]
\caption{Mean RMSE between the noise-free and noisy estimation of translation~[mm]~/~rotation~[$^\circ$]. All values are the result of averaging over five independent noise simulations of one scan. The rows and columns show the exponents $f_a$ and $f_g$ of the noise factor for the accelerometer resp. gyroscope. RMSE values for both noisy acceleration and angular velocity that are below 1 for translation and rotation are highlighted as bold.}
\label{tab-noiseSignals}
\makegapedcells
\begin{small}
\begin{tabular*}{\textwidth}{@{\extracolsep{\fill}}l|rrrrrrr@{}}
\hline
\diagbox{$f_a$}{$f_g$}        & \multicolumn{1}{c}{0} & \multicolumn{1}{c}{1} & \multicolumn{1}{c}{2} & \multicolumn{1}{c}{3} & \multicolumn{1}{c}{4} & \multicolumn{1}{c}{5} & \multicolumn{1}{c}{no noise} \\ \hline
0		& 9461 / 1.453	& 8866 / 0.146 		& 6329 / 0.016 		& 6762 / $10^{\shortminus3}$ 		& 8518 / $10^{\shortminus4}$ 				& 10389 / $2\cdot10^{\shortminus5}$ 		& 9390 / $10^{\shortminus5}$ \\
1		& 1899 / 1.404	& 883 / 0.128		& 663 / 0.012		& 754 / $10^{\shortminus3}$		& 883 / $10^{\shortminus4}$				& 666 / $2\cdot10^{\shortminus5}$				& 940 / $10^{\shortminus5}$  \\
2		& 1104 / 1.357	& 158 / 0.128		& 133 / 0.016		& 73 / $10^{\shortminus3}$		& 61 / $10^{\shortminus4}$				& 77 / $2\cdot10^{\shortminus5}$				& 81 / $10^{\shortminus5}$   \\
3		& 1173 / 1.270  & 125 / 0.122       & 20 / 0.015		& 7 / $10^{\shortminus3}$         & 9 / $10^{\shortminus4}$       			& 9 / $2\cdot10^{\shortminus5}$				& 9 / $10^{\shortminus5}$    \\
4		& 1622 / 1.645  & 190 / 0.190       & 10 / 0.014        & 1 / $10^{\shortminus3}$         & \textbf{0.943 / $\mathbf{10^{\shortminus4}}$}	& \textbf{0.985 / $\mathbf{2\cdot10^{\shortminus5}}$}	& 0.672 / $10^{\shortminus5}$    \\
5		& 108 / 1.193   & 146 / 0.178       & 10 / 0.013        & 1 / $10^{\shortminus3}$         & \textbf{0.167 / $\mathbf{10^{\shortminus4}}$}	& \textbf{0.078 / $\mathbf{2\cdot10^{\shortminus5}}$}	& 0.077 / $10^{\shortminus5}$    \\
no noise	& 148 / 1.285   & 80 / 0.129        & 12 / 0.014		& 1 / $10^{\shortminus3}$         & 0.077 / $10^{\shortminus4}$         	& 0.021 / $2\cdot10^{\shortminus5}$			& 0 / 0    \\ \hline
\end{tabular*}
\end{small}
\end{table*}
The decremental signal noise analysis shows that the RMS noise of current commercially available community IMUs would prevent a successful IMU motion compensation (Table~\ref{tab-noiseSignals}, top left).
While the resulting rotation estimate shows an average RMSE to the noise-free estimate of 1.45$^\circ$, the value of the estimated translation is considerably larger (9461\,mm).
Deviations above 1\,mm and 1$^\circ$ of the translation and rotation are expected to decrease the reconstruction quality considerably.
For noisy acceleration and angular velocity, an average RMSE value below these thresholds was only achieved if the RMS noise value was decreased by a factor of $10^{4}$ or $10^{5}$.
For this reason, and in the further analysis, the estimated motion matrices of these noise levels are used to perform a motion compensated reconstruction.
\begin{figure*}[tb]
    \centering
    \subcaptionbox*{}{\includegraphics[width=7pc]{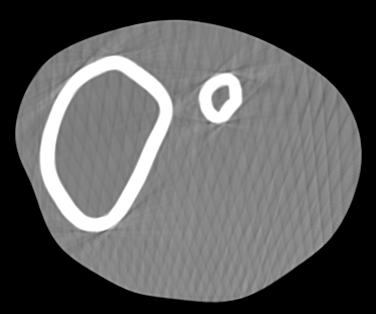}}
    \quad
    \subcaptionbox*{}{\includegraphics[width=7pc]{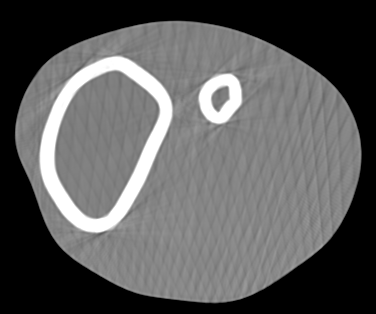}}
    \quad
    \subcaptionbox*{}{\includegraphics[width=7pc]{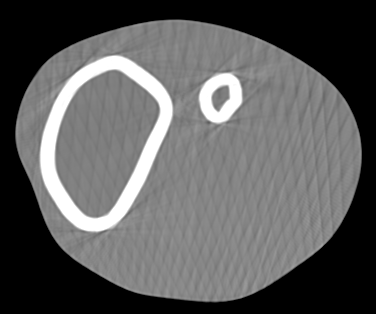}}
    \quad
    \subcaptionbox*{}{\includegraphics[width=7pc]{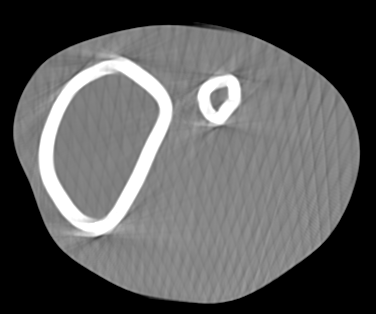}}
    \quad
    \subcaptionbox*{}{\includegraphics[width=7pc]{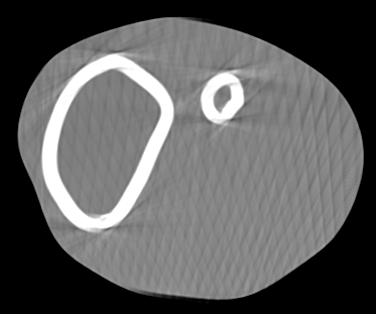}}
    \\
    \subcaptionbox*{}{\includegraphics[width=7pc]{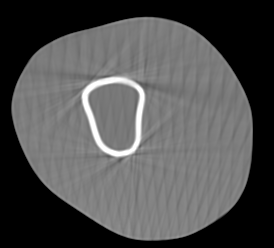}}
    \quad
    \subcaptionbox*{}{\includegraphics[width=7pc]{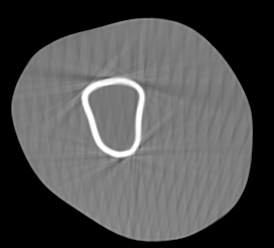}}
    \quad
    \subcaptionbox*{}{\includegraphics[width=7pc]{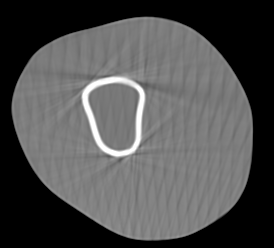}}
    \quad
    \subcaptionbox*{}{\includegraphics[width=7pc]{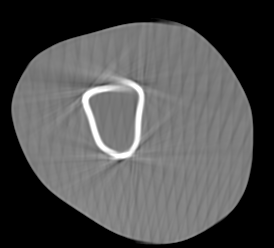}}
    \quad
    \subcaptionbox*{}{\includegraphics[width=7pc]{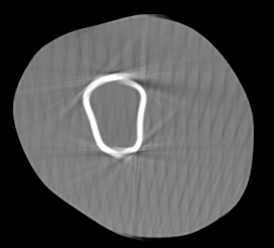}}
    \\
    \subcaptionbox{No noise}{\includegraphics[width=7pc]{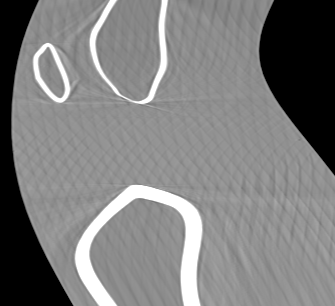}}
    \quad
    \subcaptionbox{$f_{a} = 5$, $f_{g} = 5$} {\includegraphics[width=7pc]{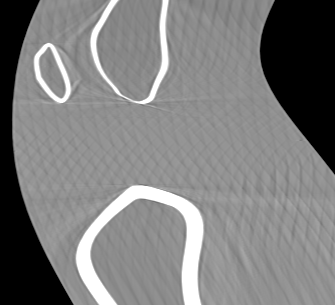}}
    \quad
    \subcaptionbox{$f_{a} = 5$, $f_{g} = 4$} {\includegraphics[width=7pc]{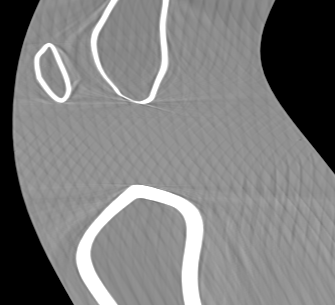}}
    \quad
    \subcaptionbox{$f_{a} = 4$, $f_{g} = 5$} {\includegraphics[width=7pc]{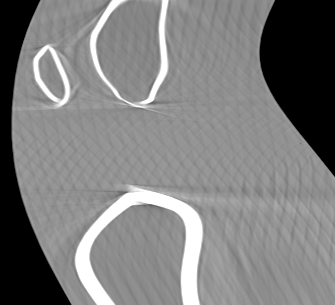}}
    \quad
    \subcaptionbox{$f_{a} = 4$, $f_{g} = 4$} {\includegraphics[width=7pc]{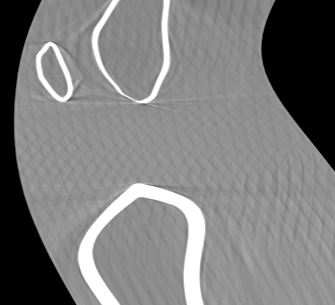}}
    \caption{Comparison of noise-free and noisy rigid IMU compensation. Rows: axial slice through shin, axial slice through thigh, sagittal slice. (a) Noise-free IMU signal, in row (b)-(e) noise is added to the simulated acceleration and angular velocity. The RMS noise value is $1.8\,mg/s^2$ resp. $0.07\,^\circ$/s divided by $10^{f_{a}}$ resp. $10^{f_{g}}$.}
    \label{fig-resultnoise}
\end{figure*}
The resultant reconstructions are shown in Fig.~\ref{fig-resultnoise}.
While the image quality for $f_a = 5$ is similar to the motion-free case, streaking artifacts are visible when $f_a = 4$, independent of $f_g = 4$ or $f_g = 5$.
\begin{table*}[tp]
\caption{Average SSIM and RMSE values with standard deviation of the rigid motion compensated reconstructions from noisy signals compared to the motion-free reconstruction over all scans. The noise factor is $10^{-f_a}$ for the simulated accelerometer resp. $10^{-f_g}$ for the gyroscope. All measures were computed on the whole reconstructed part of the leg, and on the shank resp. thigh only.}
\label{tab-noise}
\begin{small}
\begin{tabular*}{\textwidth}{@{\extracolsep{\fill}}l|lll|lll}
\hline
                 & \multicolumn{3}{c|}{SSIM}                                                              & \multicolumn{3}{c}{RMSE}                                                              \\ \hline
                 & \multicolumn{1}{c}{Whole Leg} & \multicolumn{1}{c}{Shank} & \multicolumn{1}{c|}{Thigh} & \multicolumn{1}{c}{Whole Leg} & \multicolumn{1}{c}{Shank} & \multicolumn{1}{c}{Thigh} \\ \hline
No noise        & 0.991 $\pm$ 0.004              & 0.991 $\pm$ 0.004         & 0.990 $\pm$ 0.006          & 0.017 $\pm$ 0.006             & 0.016 $\pm$ 0.006         & 0.017 $\pm$ 0.006         \\
$f_a=5,f_g=5$     & 0.989 $\pm$ 0.005              & 0.990 $\pm$ 0.005         & 0.987 $\pm$ 0.004          & 0.018 $\pm$ 0.006             & 0.018 $\pm$ 0.006         & 0.019 $\pm$ 0.006         \\
$f_a=5,f_g=4$     & 0.987 $\pm$ 0.004              & 0.988 $\pm$ 0.004         & 0.985 $\pm$ 0.004	      & 0.020 $\pm$ 0.006            & 0.020 $\pm$ 0.006         & 0.021 $\pm$ 0.007         \\
$f_a=4,f_g=5$     & 0.926 $\pm$ 0.049              & 0.929 $\pm$ 0.046         & 0.919 $\pm$ 0.054          & 0.047 $\pm$ 0.022            & 0.046 $\pm$ 0.022         & 0.048 $\pm$ 0.023         \\
$f_a=4,f_g=4$     & 0.927 $\pm$ 0.043              & 0.929 $\pm$ 0.040	     & 0.923 $\pm$ 0.052          & 0.048 $\pm$ 0.021             & 0.048 $\pm$ 0.021         & 0.047 $\pm$ 0.021         \\ \hline
\end{tabular*}
\end{small}
\end{table*}

The quantitative analysis of the noisy results in Table~\ref{tab-noise} confirms this finding: The average SSIM and RMSE values are only slightly decreased respectively increased compared with the noise-free estimation if $f_a = 5$, but deteriorate markedly when $f_a = 4$.

\section{Discussion and conclusion}
With the presented simulation study, we have shown the feasibility and limitations of using IMUs for motion compensated CT reconstruction.
While all proposed methods are capable of reducing motion artifacts in the noise-free case, our noise analysis shows that the applicability in real settings is not yet fully realizable.

The presented initialization approach based on the system geometry and the first two projection images works well under the optimal conditions of a simulation.
In a real setting, it is unlikely that the IMU will contain clearly distinguishable metal components at the IMU coordinate system and they are unlikely to be resolved using current flat panel detectors.
However, the presented approach can be applied with arbitrary four IMU points, assuming their relation to the origin and coordinate system is known.
The IMU should then be positioned such that their projections are well distinguishable in the two projection images required for initialization.

The results of all proposed IMU-based motion compensation methods are qualitatively and quantitatively equivalent, or even improved, compared with the gold standard marker-based approach that estimates a rigid motion.
For the marker-based approach, individual multiple tiny markers have to be placed successively, and need to be attached directly to the skin in order to limit soft tissue artifact.
For effective marker tracking, it should be ensured that they don't overlap in the projections.
The metal also produces artifacts in the knee region.
An advantage of our proposed methods is the need for only one or two IMUs on the leg.
Here, the only is that the components used for initialization need to be visible in the first projection images.
Since the shank and thigh are modeled as stiff segments, the sensors can be placed sufficiently far away from the knee joint in order not to cause metal artifacts that could hinder subsequent image analyses.

It is noticeable that all methods performed slightly better on the scans where subjects were asked to hold a squat of 30 degrees compared with those for the 60 degrees squat.
This is likely a result of it being more challenging to hold the same pose at a lower squat, where the motion in these cases has a higher range leading to increased error.

With the non-rigid 3D IMU approach, improved results are achieved compared with the rigid IMU approach, especially in the region of the thigh.
Although this is only a small improvement, it may have significant impact on further image analyses given the sub-millimeter range of the expected cartilage change under load \cite{Glaser2002}.
However, the simple model of three moving joint positions and an affine deformation is considerably less complex than the XCAT spline deformation during projection generation suggesting that further improvements can be achieved by using a more realistic model. 

The non-rigid 2D IMU approach provides small improvements in visual results compared with the rigid approach (Fig.~\ref{fig-resultzoom}), but the quantitative evaluation shows similar SSIM and RMSE values.
Although the non-rigid motion estimate might be more accurate, at the same time the image deformation introduces small errors, since X-rays measured at a deformed detector position would have also been attenuated by other materials.

It is notable that the noise has a larger effect on the processing of the accelerometer signal compared with that of the gyroscope signal (Table~\ref{tab-noiseSignals}).
On the one hand, the double integration performed on the acceleration leads to a quadratic error propagation.
On the other hand, the noisy gyroscope signals used for gravity removal and velocity integration introduce additional errors that are accumulated during acceleration processing.

In our study, we focus only on signal noise as one of the most severe IMU measurement errors.
In the future, similar simulations might be performed in order to determine further necessary specifications.

The noise level improvements that are required for real application are in the range of $10^{5}$ for the accelerometer and $10^{4}$ for the gyroscope.
Although recently developed accelerometers and gyroscopes achieve these low noise levels, they are designed to measure signals in the mg-range and are far too delicate for the application at hand \cite{Darvishia2019,Masu2020,Yamane2019}.
If developments continue to progress rapidly, and a robust sensor with low noise level and high measurement range is developed, our method could be applied in a real setting.

\section*{Acknowledgments}
This work was supported by the Research Training Group 1773 Heterogeneous Image Systems funded by the German Research Foundation (DFG), and by the AIT4Surgery grant funded by the Federal Ministry of Education and Research (BMBF, grant number 03INT506BA). Bjoern Eskofier gratefully acknowledges the support of the German Research Foundation (DFG) within the framework of the Heisenberg professorship programmme (grant number ES 434/8-1). The authors acknowledge funding support from NIH 5R01AR065248-03 and NIH Shared Instrument Grant No. S10 RR026714 supporting the zeego@StanfordLab.

%%Harvard
\bibliographystyle{splncs04.bst}
\bibliography{refs.bib}

\begin{thebibliography}{10}
\providecommand{\url}[1]{\texttt{#1}}
\providecommand{\urlprefix}{URL }
\providecommand{\doi}[1]{https://doi.org/#1}

\bibitem{Arden2006}
Arden, N., Nevitt, M.C.: Osteoarthritis: Epidemiology. Best Pract Res Cl Rh
  \textbf{20}(1),  3 -- 25 (2006)

\bibitem{Berger2016}
Berger, M., M{\"{u}}ller, K., Aichert, A., Unberath, M., Thies, J., Choi, J.H.,
  Fahrig, R., Maier, A.: {Marker-free motion correction in weight-bearing
  cone-beam CT of the knee joint}. Med Phys  \textbf{43}(3),  1235--48 (2016)

\bibitem{Bier2017}
Bier, B., Aichert, A., Felsner, L., Unberath, M., Levenston, M., Gold, G.,
  Fahrig, R., Maier, A.: {Epipolar Consistency Conditions for Motion Correction
  in Weight-Bearing Imaging}. In: {BVM 2017}. pp. 209--214 (2017)

\bibitem{Bier2018landmark}
Bier, B., Aschoff, K., Syben, C., Unberath, M., Levenston, M., Gold, G.,
  Fahrig, R., Maier, A.: Detecting anatomical landmarks for motion estimation
  in weight-bearing imaging of knees. In: {Machine Learning for Medical Imaging
  Reconstruction}. pp. 83--90 (2018)

\bibitem{Bier2018range}
Bier, B., Ravikumar, N., Unberath, M., Levenston, M., Gold, G., Fahrig, R.,
  Maier, A.: {Range Imaging for Motion Compensation in C-Arm Cone-Beam CT of
  Knees under Weight-Bearing Conditions}. J Imaging  \textbf{4}(1),  1--16
  (2018)

\bibitem{Bogert1996}
van~den Bogert, A.J., Read, L., Nigg, B.M.: A method for inverse dynamic
  analysis using accelerometry. J Biomech  \textbf{29}(7),  949 -- 954 (1996)

\bibitem{Bosch2020}
{Bosch Sensortec}: BMI160 - Data sheet (11 2020),
  \url{https://www.bosch-sensortec.com/products/motion-sensors/imus/bmi160.html},
  accessed: 2021-01-18

\bibitem{Choi2013}
Choi, J.H., Fahrig, R., Keil, A., Besier, T.F., Pal, S., McWalter, E.J.,
  Beaupr{\'{e}}, G.S., Maier, A.: {Fiducial marker-based correction for
  involuntary motion in weight-bearing C-arm CT scanning of knees. Part I.
  Numerical model-based optimization.} Med Phys  \textbf{40}(9),  091905--1 --
  091905--12 (2013)

\bibitem{Choi2014}
Choi, J.H., Maier, A., Keil, A., Pal, S., McWalter, E.J., Beaupr{\'{e}}, G.S.,
  Gold, G.E., Fahrig, R.: {Fiducial marker-based correction for involuntary
  motion in weight-bearing C-arm CT scanning of knees. II. Experiment.} Med
  Phys  \textbf{41}(6),  061902--1 -- 061902--16 (2014)

\bibitem{Darvishia2019}
Darvishia, A., Najafi, K.: Analysis and design of super-sensitive stacked (s3)
  resonators for low-noise pitch/roll gyroscopes. In: 2019 IEEE INERTIAL.
  pp.~1--4 (2019)

\bibitem{Desapio2017}
De~Sapio, V.: Advanced Analytical Dynamics: Theory and Applications. Cambridge
  University Press (2017)

\bibitem{Delp2007}
Delp, S.L., Anderson, F.C., Arnold, A.S., Loan, P., Habib, A., John, C.T.,
  Guendelman, E., Thelen, D.G.: Opensim: Open-source software to create and
  analyze dynamic simulations of movement. IEEE Trans Biomed Eng
  \textbf{54}(11),  1940--1950 (2007)

\bibitem{Glaser2002}
Glaser, C., Putz, R.: Functional anatomy of articular cartilage under
  compressive loading quantitative aspects of global, local and zonal reactions
  of the collagenous network with respect to the surface integrity.
  Osteoarthritis Cartilage  \textbf{10}(2),  83 -- 99 (2002)

\bibitem{Hamner2010}
Hamner, S.R., Seth, A., Delp, S.L.: Muscle contributions to propulsion and
  support during running. J Biomech  \textbf{43}(14),  2709 -- 2716 (2010)

\bibitem{Jost2016}
Jost, G., Walti, J., Mariani, L., Cattin, P.: A novel approach to navigated
  implantation of s-2 alar iliac screws using inertial measurement units. J
  Neurosurg: Spine  \textbf{24}(3),  447--453 (2016)

\bibitem{Kautz2017}
Kautz, T., Groh, B.H., Hannink, J., Jensen, U., Strubberg, H., Eskofier, B.M.:
  Activity recognition in beach volleyball using a deep convolutional neural
  network. Data Min Knowl Discov  \textbf{31}(6),  1678--1705 (2017)

\bibitem{Kok2017}
Kok, M., Hol, J.D., Sch{\"{o}}n, T.B.: {Using Inertial Sensors for Position and
  Orientation Estimation}. Foundations and Trends{\textregistered} in Signal
  Processing  \textbf{11}(1-2),  1--153 (2017)

\bibitem{Lemammer2019}
Lemammer, I., Michel, O., Ayasso, H., Zozor, S., Bernard, G.: Online mobile
  c-arm calibration using inertial sensors: a preliminary study in order to
  achieve cbct. Int J Comput Assist Radiol Surg  \textbf{15},  213–224 (2019)

\bibitem{Maier2011}
Maier, A., Choi, J.H., Keil, A., Niebler, C., Sarmiento, M., Fieselmann, A.,
  Gold, G., Delp, S., Fahrig, R.: {Analysis of Vertical and Horizontal Circular
  C-Arm Trajectories}. In: {Proc SPIE}. vol.~7961, pp. 796123--1--8 (2011)

\bibitem{Maier2013}
Maier, A., Hofmann, H., Berger, M., Fischer, P., Schwemmer, C., Wu, H.,
  M{\"{u}}ller, K., Hornegger, J., Choi, J.H., Riess, C., Keil, A., Fahrig, R.:
  {CONRAD - A software framework for cone-beam imaging in radiology}. Med Phys
  \textbf{40}(11),  111914 (2013)

\bibitem{Maier2012}
Maier, A., Hofmann, H., Schwemmer, C., Hornegger, J., Keil, A., Fahrig, R.:
  {Fast Simulation of X-ray Projections of Spline-based Surfaces using an
  Append Buffer}. Phys Med Biol  \textbf{57}(19),  6193–6210 (2012)

\bibitem{Maier2020}
Maier, J., Nitschke, M., Choi, J.H., Gold, G., Fahrig, R., Eskofier, B.M.,
  Maier, A.: Inertial measurements for motion compensation in weight-bearing
  cone-beam ct of the knee. In: MICCAI. pp. 14--23 (2020)

\bibitem{Masu2020}
Masu, K., Machida, K., Yamane, D., Ito, H., Ishihara, N., Chang, T.F.M., Sone,
  M., Shigeyama, R., Ogata, T., Miyake, Y.: Cmos-mems based microgravity sensor
  and its application. ECS Trans  \textbf{97}(5), ~91 (2020)

\bibitem{Powers2003}
Powers, C.M., Ward, S.R., Fredericson, M., Guillet, M., Shellock, F.G.:
  {Patellofemoral Kinematics During Weight-Bearing and Non-Weight-Bearing Knee
  Extension in Persons With Lateral Subluxation of the Patella: A Preliminary
  Study}. J Orthop Sports Phys Ther  \textbf{33}(11),  677--685 (2003)

\bibitem{Schaefer2006}
Schaefer, S., McPhail, T., Warren, J.: Image deformation using moving least
  squares. ACM Trans Graph  \textbf{25}(3),  533–540 (2006)

\bibitem{Segars2010}
Segars, W.P., Sturgeon, G., Mendonca, S., Grimes, J., Tsui, B.M.W.: 4d xcat
  phantom for multimodality imaging research. Med Phys  \textbf{37}(9),
  4902--4915 (2010)

\bibitem{Seth2018}
Seth, A., Hicks, J.L., Uchida, T.K., Habib, A., Dembia, C.L., Dunne, J.J., Ong,
  C.F., DeMers, M.S., Rajagopal, A., Millard, M., Hamner, S.R., Arnold, E.M.,
  Yong, J.R., Lakshmikanth, S.K., Sherman, M.A., Ku, J.P., Delp, S.L.: Opensim:
  Simulating musculoskeletal dynamics and neuromuscular control to study human
  and animal movement. PLOS Comput Biol  \textbf{14}(7),  1--20 (2018)

\bibitem{Sisniega2017}
Sisniega, A., Stayman, J.W., Yorkston, J., Siewerdsen, J.H., Zbijewski, W.:
  {Motion compensation in extremity cone-beam CT using a penalized image
  sharpness criterion}. Phys Med Biol  \textbf{62}(9),  3712--3734 (2017)

\bibitem{Thies2019}
Thies, M., Maier, J., Eskofier, B., Maier, A., Levenston, M., Gold, G., Fahrig,
  R.: Automatic orientation estimation of inertial sensors in c-arm ct
  projections. Curr Dir Biomed Eng  \textbf{5}(1),  195--198 (2019)

\bibitem{Wang2004}
Wang, Z., Bovik, A.C., Sheikh, H.R., Simoncelli, E.P.: Image quality
  assessment: from error visibility to structural similarity. IEEE Trans Image
  Process  \textbf{13}(4),  600--612 (2004)

\bibitem{Woodman2007}
Woodman, O.J.: {An introduction to inertial navigation}. Tech. Rep.
  UCAM-CL-TR-696, University of Cambridge, Computer Laboratory (2007)

\bibitem{Yamane2019}
Yamane, D., Konishi, T., Safu, T., Toshiyoshi, H., Sone, M., Machida, K., Ito,
  H., Masu, K.: A mems accelerometer for sub-mg sensing. Sensor Mater
  \textbf{31}(9),  2883--2894 (2019)

\bibitem{Zhu2007}
Zhu, Y., Gortler, S.J.: 3d deformation using moving least squares. Technical
  report: Tr-10-07, Harvard Computer Science (2007)

\end{thebibliography}
%\printbibliography

\end{document}